\documentclass[trackchanges,twocolumn]{aastex701}

\newcommand{\Mdot}{\mathrm{M}_{\odot}}
\newcommand{\Rdot}{\mathrm{R}_{\odot}}
\newcommand{\Ldot}{\mathrm{L}_{\odot}}
\newcommand{\Myr}{\mathrm{Myr}}

\begin{document}

\title{Unraveling the Origin of Unequal Mass Gravitational Wave Events: Insights from a Galactic High Mass X-ray Binary}

\author[orcid=0000-0002-8465-8090,sname='North America']{Neev Shah}
\affiliation{Steward Observatory, Department of Astronomy, The University of Arizona, 933 N. Cherry Ave., Tucson, AZ 85721, USA}
\email[show]{neevshah@arizona.edu}

\author[orcid=0000-0002-6718-9472,sname='North America']{Mathieu Renzo}
\affiliation{Steward Observatory, Department of Astronomy, The University of Arizona, 933 N. Cherry Ave., Tucson, AZ 85721, USA}
\email{mrenzo@arizona.edu}

\author[orcid=0000-0002-8134-4854,sname='North America']{Koushik Sen}
\affiliation{Steward Observatory, Department of Astronomy, The University of Arizona, 933 N. Cherry Ave., Tucson, AZ 85721, USA}
\email{ksen@arizona.edu}

\author[orcid=0000-0002-2215-1841,sname='North America']{Aldana Grichener}
\affiliation{Steward Observatory, Department of Astronomy, The University of Arizona, 933 N. Cherry Ave., Tucson, AZ 85721, USA}
\email{agrichener@arizona.edu}

\author[orcid=0000-0001-5228-6598,sname='North America']{Katelyn Breivik}
\affiliation{McWilliams Center for Cosmology and Astrophysics, Department of Physics, Carnegie Mellon University, Pittsburgh, PA 15213, USA}
\email{}

\begin{abstract}

The catalog of Gravitational Wave (GW) events is rapidly growing, providing key insights into the evolution of massive binaries and compact object formation. However, a key challenge is to explain the origin of exceptional events such as GW190814, among the most asymmetric mass-ratio mergers to date ($q\approx 0.1$). We show that it shares an evolutionary pathway with the most unequal mass Galactic High Mass X-ray Binary (HMXB) 4U 1700-37/ HD 153919. We demonstrate this unique connection by utilizing a rich set of existing observational constraints for the HMXB and compute detailed binary evolution models to explain its formation history. We find that conservative mass transfer, along with a directed natal kick are essential to explain its current state. We show that this system is unlikely to form a GW source due to a failed Common Envelope (CE) phase in the future, in agreement with previous work. With additional models, we show that a similar pathway naturally forms GW190814-like events, provided the first phase of mass transfer remains conservative, and the first-born (lower mass) compact object receives a large natal kick ($\gtrsim 100\,\mathrm{km/s}$) for the subsequent CE phase to be successful and form a asymmetric mass-ratio GW source. Anchored by the number of analogous Galactic HMXBs, we estimate
rates for such GW events, which broadly agree with their observed rate. Our work demonstrates a unified formation pathway for highly asymmetric mass-ratio HMXBs and GW events. Moreover, it highlights the critical role of finding and characterizing local analogs in different evolutionary phases, and using them as a bridge to understand the origin of GW sources, especially the outliers like GW190814.

\end{abstract}


\section{Introduction}

Massive stars ($> 8-10\,\Mdot$) end their lives in energetic explosions or implosions, often leaving behind a compact remnant like a black hole (BH) or neutron star (NS) \citep[e.g,][]{2021Natur.589...29B, 2025arXiv250214836J,2025NewA..12102453S, 2026enap....2..639J}. Most massive stars live in binaries and higher order systems \citep[e.g,][]{2014ApJS..213...34K, 2017ApJS..230...15M, 2023ASPC..534..275O}, a large fraction of which are in orbits close enough that they will likely interact at some point over their lifetimes \citep{2012Sci...337..444S, 2025NatAs...9.1337S}. Such systems are thought to be the progenitors of high-energy phenomena such as X-ray binaries (XRBs) \citep[e.g,][]{2006csxs.book..623T}, stripped envelope supernovae \citep[e.g,][]{2011MNRAS.412.1522S,2016MNRAS.457..328L}, gravitational wave (GW) sources \citep[e.g,][]{2018MNRAS.481.4009V, 2017ApJ...846..170T, 2022PhR...955....1M}, gamma-ray bursts (GRB) \citep[e.g,][]{1993ApJ...405..273W, 2007A&A...465L..29C}, and various other transients \citep[e.g,][]{2012ApJ...752L...2C,2022ApJ...932...84M, 2022RAA....22e5010S, 2023MNRAS.523.6041G}.

\begin{figure*}
    \centering
    \includegraphics[width=\textwidth]{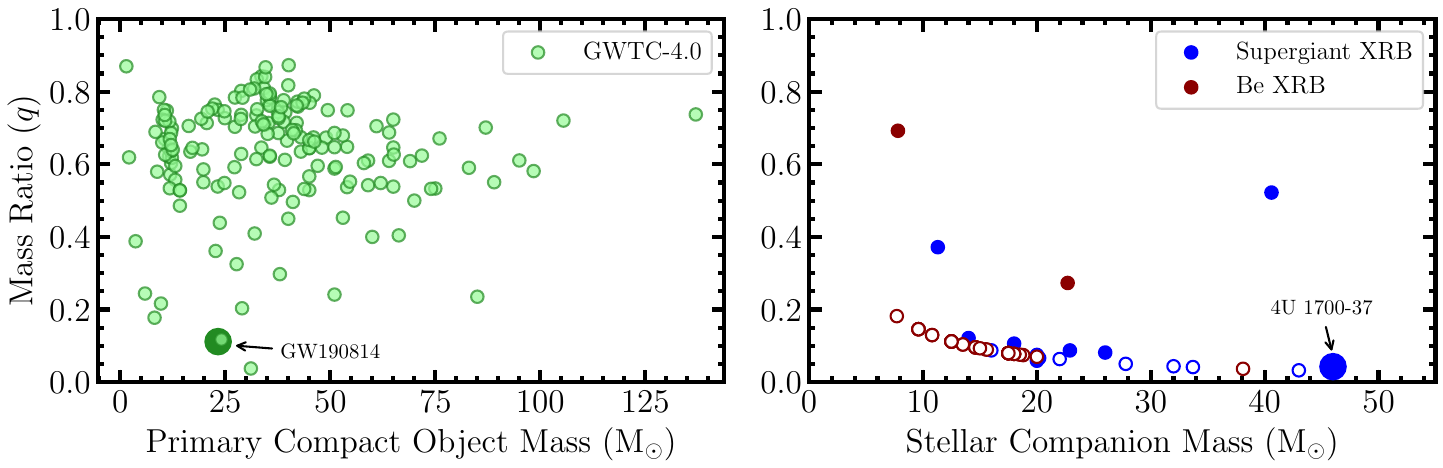}
    \caption{GW190814 and the HMXB 4U 1700-37/ HD 153919 represent extremes in their respective populations. GW190814 has one of the most asymmetric mass-ratio among GW events, while the HMXB 4U 1700-37 contains the most massive companion star among known HMXBs. \textit{Left panel:} The mass of the primary (heavier) compact object and the mass-ratio of the system for all the confident GW events in the cumulative GWTC 4.0 catalog \citep{2025arXiv250818079T}. \textit{Right panel:} The mass of the companion star and the mass-ratio of the system for all the HMXBs in the catalog of \cite{2023A&A...671A.149F}. Blue circles denote Supergiant XRBs, while red circles represent Be XRBs. Open circles correspond to systems that are cataloged as having a NS in them, but do not have a mass estimate, for which we set their mass to be $1.4\,\Mdot$. Due to the heterogeneous datasets and methods used to estimate masses, and for illustration purposes, we only report the central values here and not their measurement errors. GW190814 and 4U 1700-37 are highlighted with larger circles in both panels. The goal of this work is to demonstrate an evolutionary link between the formation of both these systems.}
    \label{fig:gw-xrb-catalog}
\end{figure*}

Although most massive binaries break at the formation of the first compact object \citep[e.g,][]{2019A&A...624A..66R}, in a minority of cases, a natural outcome is the presence of compact object(s) in binaries. In the local Universe, these systems have been observed for decades as X-ray binaries (see \citealp{2006ARA&A..44...49R} for a review). More recently, astrometry with \emph{Gaia}, and other radial velocity surveys have unraveled a small but growing population of non-interacting binaries that contain a compact object such as a BH \citep[e.g,][]{2023MNRAS.518.1057E,2023MNRAS.521.4323E,2023AJ....166....6C,2024A&A...686L...2G, 2022NatAs...6.1085S} or NS \citep[e.g,][]{2024OJAp....7E..58E}. Additionally, a handful of binary neutron star (BNS) systems have been observed in the Galaxy through pulsar observations \citep[e.g,][]{2003Natur.426..531B, 2005AJ....129.1993M}.

Double compact binaries are strong sources of GWs, especially close to merger. GWs from merging binary black holes (BBH) were directly detected for the first time in 2015 \citep{2016PhRvL.116f1102A}. The sample has rapidly grown since, with over $200$ mergers in the latest catalog, GWTC-4.0 \citep{2025arXiv250818082T,2025arXiv250818083T}.

In the catalog are a few exceptional events, such as one of the most\footnote{Although GW191219\_163120 has a mass-ratio lower than GW190814, it has a lower SNR and the LVK reports its parameters as prone to systematic uncertainties due to the lack of calibrated waveform models in that region of parameter space \citep{2023PhRvX..13d1039A}.} asymmetric mass-ratio\footnote{In this work, unless otherwise specified, we define the mass-ratio as $q = m_2/m_1$, where $m_2 \leq m_1$.} merger, GW190814, between a compact object of $\sim 2.6\,\Mdot$, and a $\sim 23\,\Mdot$ BH \citep{2020ApJ...896L..44A}, see Fig. \ref{fig:gw-xrb-catalog}. The highly asymmetric mass-ratio of this event, and the presence of a compact object straddling the ``lower mass-gap" between the heaviest neutron stars and lightest black holes (e.g, \citealp{2010ApJ...725.1918O, 2011ApJ...741..103F}, although see \citealp{2024ApJ...970L..34A}) has been challenging to interpret for most formation channels. Various attempts have been made to explain its formation, such as through isolated binary evolution \citep[e.g,][]{2020ApJ...899L...1Z, 2021MNRAS.500.1380M, 2025arXiv251116648M}, a triple scenario where the $2.6\,\Mdot$ compact object is the remnant of a previous BNS merger \citep{2021MNRAS.500.1817L}, and through dynamical interactions in young star clusters \citep[e.g,][]{2021ApJ...908L..38A}.

Due to the unknown BH or NS nature of the lower mass compact object in GW190814, it is often categorized as an outlier to the underlying population (e.g, \citealp{2021ApJ...913L...7A,2022ApJ...926...34E}, however, see also \citealp{2022ApJ...931..108F}). Outlier events challenge conventional channels for the formation of such systems. Including or excluding them from analyses can significantly bias the inferred population parameters \citep[e.g,][]{2025arXiv250814159C}. Therefore, it is important to carefully characterize and study outlier events to understand whether they are true outliers that hint towards an exotic formation scenario, or just rare outcomes of an otherwise standard evolutionary pathway.

GW events are a rare product of massive binary evolution (see \citealp{2022LRR....25....1M} for a review on rates). This makes it challenging to reconstruct their past evolutionary history on an individual event-by-event basis just from the estimated masses and/ or spins of the compact objects \citep[e.g,][]{2023ApJ...950..181W}. Instead, most studies search for signatures of different formation channels in the overall population properties, such as the mass, redshift and spin distributions, and possible correlations among them (\citealp{2022ApJ...940..184V}, see \citealp{2024arXiv241019145C} for a review).

However, another way of understanding the formation of GW sources is by finding and characterizing their analogs in earlier evolutionary stages. Such systems are often observed and well studied in the Milky Way and nearby galaxies. They are important anchors in the pathway to forming a GW source, and provide constraints on uncertainties in massive binary evolution, such as the details of the mass transfer (MT) process \citep[e.g,][]{2025arXiv251115347S}, the explodability and natal kick received to compact remnants \citep[e.g,][]{2024PhRvL.132s1403V}, and uncertainties in common envelope evolution. Such uncertainties propagate in forming a GW source, making their astrophysical interpretation a challenging task.

Utilizing local analogs can be particularly useful for understanding the formation of outlier events such as GW190814, which may not fit well with the rest of the GW population. A key intermediate stage in the pathway to becoming a gravitational wave source is a potential XRB phase \citep[e.g,][]{2021A&A...652A.138S}. Although few presently observed XRBs in the local Universe will result in merging compact binaries in the future \citep[e.g,][]{2022ApJ...929L..26F,2022ApJ...938L..19G}, any compact binary merger originating from isolated binary evolution will experience a (brief) XRB phase during its evolution \citep[e.g,][]{2020A&A...638A..39L}. In Fig. \ref{fig:gw-xrb-catalog}, we show the mass vs mass-ratio for the known population of GW events and Galactic HMXBs.

Given the extreme mass-ratio of GW190814, a potential local analog for it is the runaway HMXB\footnote{4U 1700-37 refers to the X-ray source, while HD 153919 denotes the optical counterpart at the same location. In this work, we refer to HMXB 4U 1700-37 as the binary system, unless otherwise specified.} 4U 1700-37/ HD 153919 \citep{2015A&A...577A.130F}. One of the first X-ray sources discovered in the 1970s \citep{1973ApJ...181L..43J}, the $3.41$ day period binary consists of an O6.5 Iaf+ star HD 153919 \citep{1973AJ.....78.1067W}. With an estimated mass of $40-60\,\Mdot$ \citep[e.g,][]{2003A&A...407..685H, 2015A&A...577A.130F}, it is one of the most massive stars in known XRBs (see Fig. \ref{fig:gw-xrb-catalog}). The compact object in this system is equally interesting, with initial estimates putting its mass at $\approx 2.5\,\Mdot$ \citep{2002A&A...392..909C}. However, recent observations hint that its mass may be lower at $\approx 2\,\Mdot$ \citep{2015A&A...577A.130F}, but its NS or BH nature remains unclear \citep[e.g,][]{2026arXiv260204622W}. Using Hipparcos data, \cite{2001A&A...370..170A} identified NGC 6231 as the potential parent cluster of this system. This was confirmed by \cite{2021A&A...655A..31V} using \emph{Gaia} EDR3, and verified by \cite{2022MNRAS.511.4123H} using the full \emph{Gaia} DR3. They find that the HMXB was ejected from NGC 6231 about $2\,\Myr$ ago, presumably due to the SN explosion that formed the compact object. The young age ($< 10\,\Myr$) of NGC 6231 \citep[e.g,][]{2007MNRAS.377..945S} implies that the progenitor star of the compact object had an initial mass $>\,30\,\Mdot$ \citep{2001A&A...370..170A}.

The short period of the binary, its highly asymmetric mass-ratio, and young age have been challenging to interpret. \cite{2021A&A...655A..31V} proposed the system to have formed from a massive binary that underwent mass transfer while the more massive (donor) star was still on the Main Sequence (MS)\footnote{Such a scenario was originally proposed by \cite{2008ApJ...685..400B} to explain the formation of an isolated neutron star in the young Westerlund 1 cluster.}, termed as Case A MT \citep{1967ZA.....65..251K}. They also hypothesized that the compact remnant received a natal kick in a favorable direction that shrinks the orbit. The presence of a compact object in the lower mass-gap, and the highly asymmetric mass-ratio of this system makes it an interesting case study for understanding the formation of asymmetric mass-ratio systems such as GW190814, as they may share a common evolutionary pathway.

Our goal in this work is two-fold. In Section \ref{sec:monte-carlo}, we first utilize the rich set of observational constraints for the HMXB 4U 1700-37/ HD 153919, and using Monte Carlo simulations reconstruct the properties of the binary prior to the first (and so far the only) SN in the system. With these constraints, in Section \ref{xrb} we compute detailed binary evolution models with \textsc{MESA} to chart its past evolutionary history and likely progenitor system. We evaluate the future evolution of the HMXB in Section \ref{xrb-future}. In Section \ref{sec:gw190814}, we compute additional detailed binary evolution models to explore the formation of GW190814-like systems. We demonstrate the conditions in which such a system can form, and its similarities and differences with the HMXB described earlier. We also provide estimates for the rates of such events in Section \ref{rates}. In Sections \ref{sec:discussion} and \ref{sec:conclusion}, we discuss certain implications and caveats for our models, and summarise our results.

\section{Reconstruction of the HMXB 4U 1700-37 before supernova} \label{sec:monte-carlo}

In this section, we use Monte Carlo simulations to infer the pre-SN properties of the runaway HMXB 4U 1700-37/ HD 153919. We utilize several observational constraints such as the mass of the companion star and the compact object, the current orbital period, and its systemic velocity with respect to its parent cluster NGC 6231.

The occurence of a SN in a binary can have a significant effect on its orbital properties \citep[e.g,][]{1961BAN....15..291B, 1994astro.ph.12023B,2025OJAp....8E..85W}. There can be diverse outcomes depending on the amount of mass lost in the explosion. The momentum carried by the SN ejecta can change the binary orbit (referred to as the ``Blauuw kick'', \citealp{1961BAN....15..265B}). Due to asymmetries in the explosion and/or neutrino emission, the compact remnants can also receive a natal kick (\cite{1975Natur.253..698K}, see \cite{2025NewAR.10101734P} for a review). The strength and directions of natal kicks, and its dependence on the nature of the compact remnant (NS or BH) are not well known (e.g, \citealp{2024PhRvL.132s1403V, 2025PASP..137c4203N,2025ApJ...989L...8D, 2025arXiv250508857V}). The combination of mass loss and natal kicks can lead to several observational consequences such as binary disruption \citep[e.g,][]{1997A&A...318..812D,2019A&A...624A..66R}, and if it remains bound, a change in the orbital period, eccentricity and even a systemic velocity to the binary as a whole.

\subsection{Methods} \label{sec:monte-carlo-method}

With the help of observational constraints on the parameters of a post-SN binary, we can infer its pre-SN parameters. We follow the setup in \cite{2024OJAp....7E..38E}, which in turn uses \cite{1994astro.ph.12023B} for calculating the outcomes of a SN in a binary. These calculations implicitly assume that the mass lost in a SN explosion is instantaneously removed from the system, given the very high ejecta speeds in a SN in comparison to the orbital velocity.

We start by creating an initial population of pre-SN binaries, where we sample over the pre-SN masses and orbital periods. We assume that the pre-SN binary orbit is circular. This is justified as we model short period binaries that undergo mass transfer, where tides are expected to have circularized the orbit (however, see \cite{2009MNRAS.400L..20E, 2025ApJ...983...39R,2025arXiv250905243P} for studies on eccentric MT). The post-SN binary properties depend on the pre-SN binary parameters and the details of the SN explosion. The relevant parameters are the amount of mass lost in the SN, and the speed and direction of the compact object natal kick. We neglect any change in the mass of the companion star during the SN \citep{1998A&A...330.1047T,2018ApJ...864..119H,2021MNRAS.505.2485O}. This specifies all the relevant parameters needed to generate the post-SN population from the pre-SN binary population.

The mass of the progenitor star that forms the compact object is denoted by $m_{1,\mathrm{pre-SN}}$, while its companion has a mass $m_2$. We denote the pre-SN orbital period as $P_{\mathrm{pre-SN}}$. We generate a population of $10^7$ binaries with $m_2$ sampled uniformly between $40-50\,\Mdot$, which roughly spans the estimated mass of the visible companion today \citep{2021A&A...655A..31V}. We fix the post-SN mass of the compact object ($m_{1,\mathrm{post-SN}}$) to be $2 \, \Mdot$, and sample the mass lost in the SN explosion ($\Delta m$) uniformly between $0-20 \, \Mdot$. Therefore, the pre-SN mass of the progenitor of the compact object as $m_{1,\mathrm{pre-SN}} = 2\, \Mdot + \Delta m$. We assume that the mass of the compact object immediately post-SN is similar to its measured mass today, as compact objects in HMXBs are not expected to gain substantial mass due to the short timescale of the HMXB phase \citep[e.g,][]{2012ApJ...747..111W, 2023pbse.book.....T}.

We sample the pre-SN period of the binary uniformly between $0-30\, \mathrm{days}$. Although we allow for extremely short periods pre-SN, they are not realistic as the stars will not fit inside a very small orbit. We sample the natal kicks to the compact object isotropically in direction, with magnitudes ($v_{\mathrm{kick}}$) from a lognormal distribution with $\mu = 5.60$ and $\sigma = 0.68$ \citep{2025ApJ...989L...8D}.

Having defined the pre-SN binary population ($m_{1,\mathrm{pre-SN}}\,, m_2\,, P$) and the relevant parameters for the SN explosion ($\Delta m\,, v_{\mathrm{kick}}$)
, we can now compute the properties of the post-SN binary (or disrupted) population. From the post-SN population, we select for systems that satisfy the observational constraints for the HMXB 4U 1700-37, (i.e,) they are bound ($e<1$) and have gained a systemic velocity ($v_{\mathrm{sys}}$) between $50-80 \, \mathrm{km/s}$ \citep{2021A&A...655A..31V}. We assume that most of its systemic velocity with respect to its birth cluster can be attributed to its post-SN systemic velocity, neglecting its evolution over time (during the runaway phase) and the impact of the velocity dispersion of the cluster \citep{2021A&A...645L..10R}. However, the details of this depend on accurately modeling the effects of the cluster or galactic potential \citep{2025AJ....170..192W}.

The presen day orbital period of the XRB is $ \sim 3.41 \, \mathrm{days}$ \citep{2016MNRAS.461..816I}, but it is possible that the period immediately post-SN period was different. It could have evolved to its current state due to wind mass loss from the companion star (that widens the orbit) and tidal effects (which may either shrink or widen the orbit, e.g, \citealp{2024A&A...684A..35B}). To account for this, we relax our post-SN period constraints and select for systems that have a post-SN orbital period between $2-5 \, \mathrm{days}$. We choose a lower limit of $2 \, \mathrm{days}$, as shorter periods imply that the companion star would have filled its Roche Lobe after the SN, which is not plausible given its current evolutionary state.

We do not impose any constraints on eccentricty, other than that it is bound post-SN, as the current eccentricity (if any) of the binary may not be representative of what it was immediately post-SN due to tidal circularization.

\begin{figure}[b]
    \centering
    \includegraphics[width=0.47\textwidth]{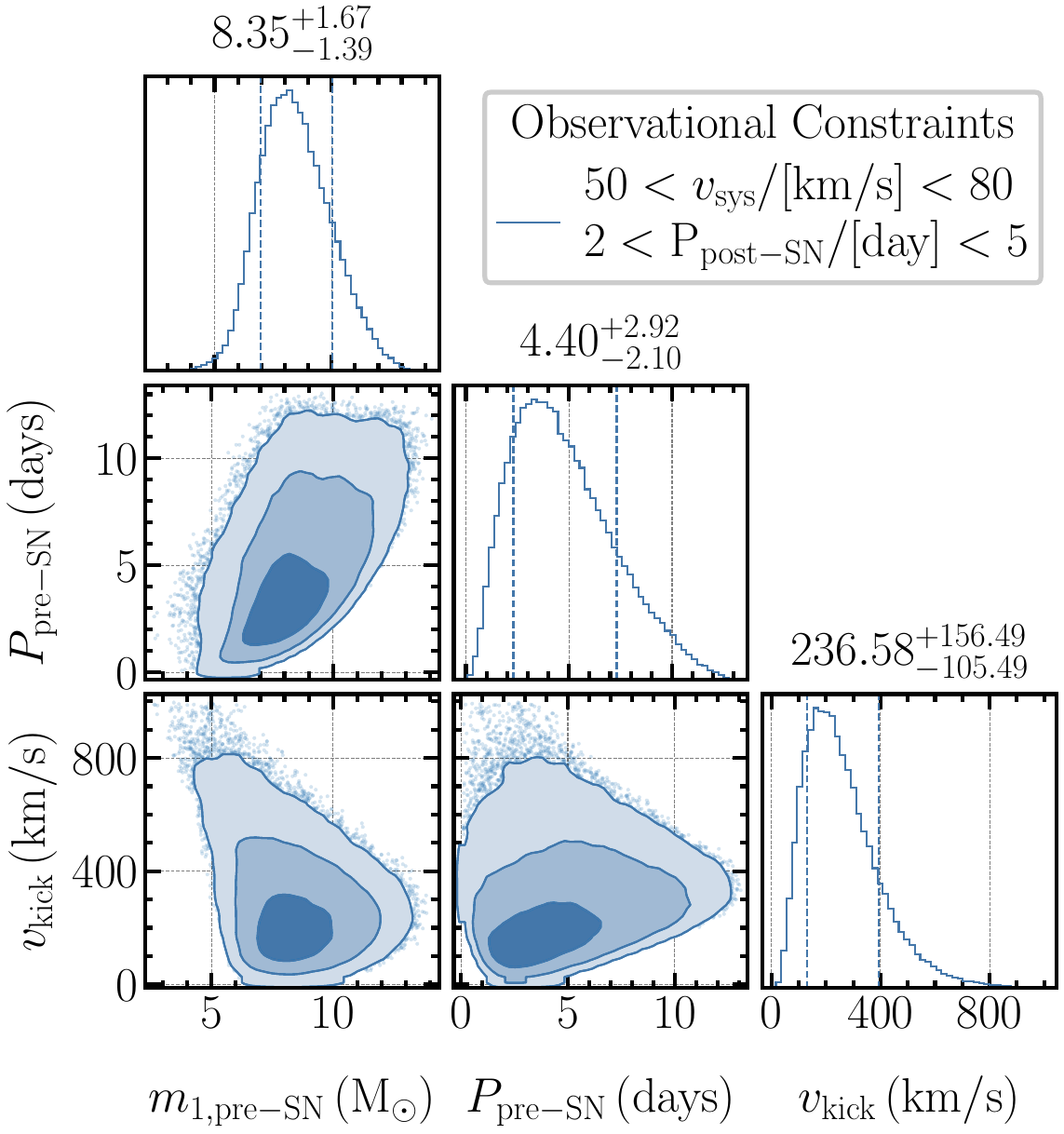}
    \caption{Reconstruction of the properties of the HMXB 4U 1700-37 pre-SN. The corner plot shows distributions of the pre-SN mass, pre-SN period and natal kick strength of the binary population that satisfy the observational constraints of the HMXB. Diagonal panels show their 1D marginal distributions (with median and $1\sigma$ uncertainties reported on the top), while other panels show correlations between different parameters, with darker colors representing regions of higher density. The goal of our binary evolution models is to match these pre-SN constraints.}
    \label{fig:xrb_monte_carlo}
\end{figure}

\subsection{Results} \label{sec:monte-carlo-results}

In Fig. \ref{fig:xrb_monte_carlo}, we show a corner plot of the pre-SN mass of the compact object progenitor, the natal kick imparted to its remnant, and the pre-SN orbital period of the systems that satisfy the imposed observational constraints. The shape of the probability distribution depends on the assumed pre-SN population. In this scenario, we uniformly explore the entire parameter space of interest that is physically relevant to this system given its observational constraints (listed in the legend in Fig. \ref{fig:xrb_monte_carlo}).

We find that the pre-SN mass of the progenitor of the compact object must have been between $4-13\,\Mdot$ to explain its current orbital period and systemic velocity. Due to the short post-SN orbital period seen today, we also find that the pre-SN orbital period is constrained to be less than $13$ days. Although we do not impose any constraints on the pre-SN orbital period, it is unlikely to have been very small ($\lesssim 1-2$ days) as the two stars would not have fit inside a very small pre-SN orbit. We also find that large pre-SN periods ($>5$ days) require large natal kicks ($>100 \, \mathrm{km/s}$), as it is not possible to shrink a binary's post-SN period just through mass loss during the SN. This is important because the first phase of MT in binaries, if stable, leads to mass-ratio reversal and orbital widening in most scenarios \citep{2019A&A...624A..66R}. We also know that this system originated in a young cluster with a high turn-off mass $\geq 30\,\Mdot$. Since our Monte Carlo simulations suggest that the progenitor of the compact object had a pre-SN mass $\leq 13\,\Mdot$, it must have lost a large amount of mass. This strongly suggests that this system likely went through a phase of MT in the past (from the progenitor of the compact object onto the current visible star), which would have widened its orbit. Therefore, the pre-SN period was likely large enough that a natal kick is necessarily needed to shrink the orbit post-SN to its current state \citep{2021A&A...655A..31V}. Thus, although our Monte Carlo simulations allow for pre-SN periods smaller than $\approx 5\,\mathrm{days}$, we do not expect such solutions to be physically plausible.

\section{Evolutionary History of the HMXB 4U 1700-37/ HD 153919} \label{xrb}

We utilize the pre-SN constraints that we obtained from our Monte Carlo simulations to reconstruct the past evolutionary history of the HMXB 4U-1700-37/ HD 153919 with detailed binary evolution models, and describe our setup below.

\subsection{Methods} \label{sec:mesa_xrb-methods}

We simulate the evolution of massive binaries using the stellar
evolution code \textsc{MESA}
\citep[][\texttt{r24.08.01}]{2011ApJS..192....3P,2013ApJS..208....4P,2015ApJS..220...15P,2018ApJS..234...34P,2019ApJS..243...10P,2023ApJS..265...15J}. Our inlists and output data are publicly available\footnote{ We also include scripts to reproduce all the figures in this work at \href{https://doi.org/10.5281/zenodo.18615967}{doi:10.5281/zenodo.18615967}} at \href{https://doi.org/10.5281/zenodo.18615967}{doi:10.5281/zenodo.18615967} and we summarize relevant parameters and assumptions below. We also perform resolution tests, and describe them in Appendix \ref{resolution}.

We use the Ledoux criterion \citep{1947ApJ...105..305L} to find
convective zones, and assume a mixing length parameter
$\alpha_{\mathrm{MLT}}=2.0$. We include semiconvection
($\alpha_{\mathrm{sc}}=1$) \citep{1983A&A...126..207L}, and
thermohaline mixing ($\alpha_{\mathrm{th}}=1$)
\citep{1980A&A....91..175K}. To account for convective boundary overshooting above the core, we use the exponential model from \cite{2000A&A...360..952H} with free parameters $(f,f_0) = (4.25 \times 10^{-2}, 10^{-3})$ \citep{2018ApJ...859..100C}, which broadly reproduces the width of the Main Sequence (MS) for a single $16 \, \Mdot$ star \citep{2011A&A...530A.115B}. We do not account for convective boundary over/undershooting for off-center convective layers. To aid in regions at the Eddington limit where convection is inefficient, we utilize the local implicit enhancement of the convective flux in superadiabatic regions \citep[\texttt{use\_superad\_reduction} from][]{2023ApJS..265...15J}.

To limit the computational difficulties in modeling massive
interacting binaries, we include rotation only in the accretor, while
we fix the donor to be non-rotating\footnote{However, see \cite{2023MNRAS.524.4315H} for impacts of the donor's rotation on mass transfer dynamics, although such effects are not explicitly included in 1D stellar evolution codes such as \textsc{MESA}.}. The accreted material carries
angular momentum which can spin-up the star, which is observed in
accretor stars, for example, a fraction of runaways
\citep{1993ASPC...35..207B} such as $\zeta\, \mathrm{Ophiuchi}$
\citep{2018AN....339...46Z, 2021ApJ...923..277R}, in Be stars
\citep{2026enap....2..430R} and blue lurkers or stragglers
\citep{2019ApJ...881...47L}. We treat rotation in the ``shellular''
approximation \citep{1992A&A...265..115Z,2012A&A...537A.146E}, where
the angular velocities $\omega$ are constant on isobaric surfaces,
effectively assuming strong angular momentum transport in the
horizontal direction due to turbulence. We start our accretor models
with a $\Omega/\Omega_{\mathrm{crit}} = 0$ at Zero-Age Main Sequence
(ZAMS), but this can change due to the gain/ loss of angular momentum through tides, winds or mass transfer. We also include the rotational
enhancement of the wind mass loss \citep{1998A&A...329..551L} to prevent the accretor from reaching critical rotation and regulate the
mass-transfer efficiency \cite[see below and also][]{2025ApJ...990L..51L, 2025arXiv251115347S, 2026arXiv260206259X}. We include several
rotational mixing processes in a diffusion approximation, such as
Eddington-Sweet circulation \citep{1950MNRAS.110..548S}, secular and
dynamic shear instabilities, and the Goldreich-Schubert-Fricke (GSF)
instability (see \citealp{2000ApJ...528..368H} for a review of these
processes). We assume a Spruit-Taylor dynamo
\citep{2002A&A...381..923S} for treating the transport of angular
momentum, and choose the same parameters as in
\cite{2000ApJ...528..368H}.

Stellar winds are a critical ingredient of massive star evolution, as they can modify their internal structure \citep[e.g,][]{2017A&A...603A.118R, 2026arXiv260102263R} and also affect the orbital evolution in binaries. For stars with a surface effective temperature $T_\mathrm{{eff}} > 11000\,\mathrm{K}$, we utilize the mass loss prescriptions from \cite{2023A&A...676A.109B}, unless the surface hydrogen mass fraction falls below $0.4$, in which case we use the optically thick Wolf-Rayet (WR) wind mass loss rates from \cite{2000A&A...360..227N}. For cooler stars with a $\mathrm{T_{eff}} < 10000\,\mathrm{K}$, we use the mass-loss rates from \cite{2024A&A...681A..17D} but do not allow it to exceed $10^{-6}\,\Mdot/\mathrm{yr}$. We interpolate linearly between the hot and cool winds for the temperature ranges $10000-11000\,\mathrm{K}$.

We treat mass transfer in a binary using the scheme described
in \cite{1990A&A...236..385K}. We assume that mass transfer is fully
conservative until the accretor reaches a maximum
$\Omega/\Omega_{\mathrm{crit}} = 0.9$, after which the mass-transfer
efficiency depends on the rotationally enhanced wind mass loss. We
assume that the specific angular momentum and entropy of the
transferred mass is equal to the corresponding values at the surface
of the accretor, and the chemical composition is set by the
stratification of the donor. The transferred mass that is not accreted
leaves the binary system carrying with it the specific angular
momentum of the orbital motion of the accretor
\citep{1997A&A...327..620S,2017MNRAS.471.4256V}. We also
include tidal effects \citep{1977A&A....57..383Z,
1981A&A....99..126H,2002MNRAS.329..897H}, which can have a significant
impact on the spins and orbital evolution of close binaries.

\subsection{Results}\label{sec:mesa-xrb-results}

Based on the results of our Monte Carlo simulations in Section \ref{sec:monte-carlo-results}, we compute binary evolution models at solar metallicity ($Z_\odot = 0.0146$, \citealp{2009ARA&A..47..481A}) that can reproduce i) the pre-SN mass of the progenitor of the compact object ii) the estimated mass of the companion star in the HMXB today and iii) the pre-SN orbital period of the binary. To illustrate the evolution and structure of our binary models,
we focus our presentation on a fiducial model that can
explain the past evolutionary history of the HMXB 4U 1700-37. This
corresponds to a binary of initial masses $40\,\Mdot$ and $28\,\Mdot$
in a tight $3\,\mathrm{day}$ orbit. In Appendix \ref{grid}, we show a small grid with variations in the initial parameters.

\begin{figure}[htbp]
    \centering
    \includegraphics[width=0.45\textwidth]{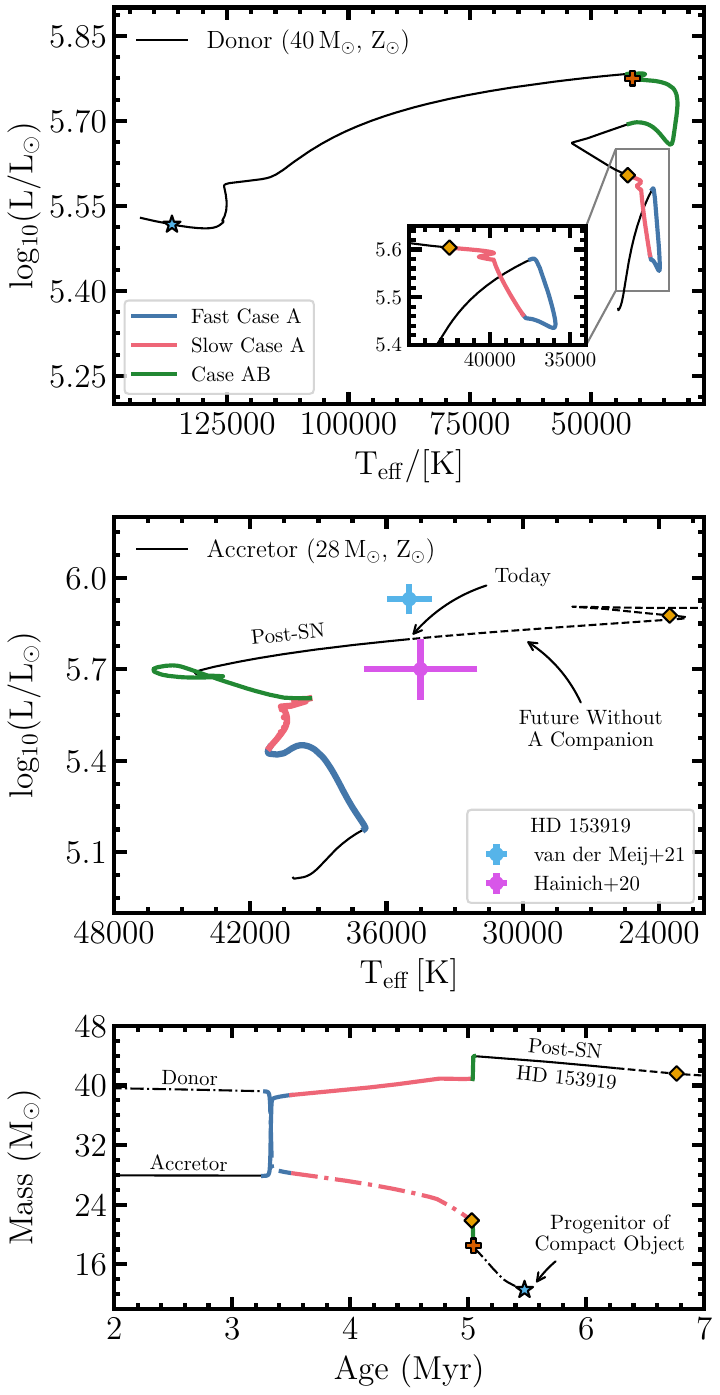}
    \caption{The evolution of the donor and accretor in our fiducial model for forming the HMXB 4U 1700-37/ HD 153919 at $Z = Z_\odot = 0.0146$. Their initial masses are $40\,\Mdot$ and $28\,\Mdot$, respectively, and the initial orbital period is $3\,\mathrm{days}$. \textit{Top panel:} Evolution of the donor on the HR diagram. Blue  and red curves highlight fast and slow Case A mass transfer, respectively, while green shows Case AB mass transfer. The diamond, plus, and star markers correspond to the Terminal Age Main Sequence (TAMS), core-helium ignition, and core-helium depletion, respectively.
    The inset panel zooms into the evolution of the donor during Case A mass transfer. \textit{Middle Panel:} Evolution of the accretor on the HR diagram, representing our model for HD 153919. The purple and blue markers with errorbars show the observed location of HD 153919 from \cite{2020A&A...634A..49H} and \cite{2021A&A...655A..31V}, respectively. \textit{Bottom panel:} The evolution of the mass of the donor (dash-dot) and accretor (solid) over time.}
    \label{fig:xrb_fiducial_hr}
\end{figure}

In Fig. \ref{fig:xrb_fiducial_hr}, we show the evolution of the donor (top) and accretom (middle) on separate HR diagrams. We also show the evolution of their masses in the bottom panel. Due to the short initial period, the donor overflows its Roche Lobe after $3.26\,\mathrm{Myr}$, during the Main Sequence (MS) itself. This leads to a phase of mass transfer from the donor to the accretor, referred to as Case A \citep{1967ZA.....65..251K}. This consists of an initial thermal timescale mass transfer ($\sim 0.2\, \mathrm{Myr}$), known as fast Case A (highlighted in blue in Fig. \ref{fig:xrb_fiducial_hr}). As the donor regains thermal equilibrium, it slowly grows in size on a nuclear timescale, which leads to a steady period of mass transfer referred to as slow Case A (highlighted in red in Fig. \ref{fig:xrb_fiducial_hr}). This phase ends around $\sim 5\,\mathrm{Myr}$, which roughly corresponds to when the donor has finished its supply of hydrogen in its core. It starts contracting on a thermal timescale, ending RLOF. At this stage, the donor has lost around $\sim 18\,\Mdot$ and has a mass of $\sim 22\,\Mdot$. Due to the stripping of its envelope during MT, the surface has an elevated helium mass fraction ($Y_{\mathrm{surf}} = 0.67$) and is hot with a $T_{\mathrm{eff}} \sim 42500\,\mathrm{K}$. The mass of the helium core at this point is almost $17.2\,\Mdot$.

On the other hand, the accretor (still on the MS) has grown from an initial $28\,\Mdot$ to $\approx 40.9\,\Mdot$. This corresponds to an average mass transfer efficiency $\left\langle\beta_{\mathrm{RLOF}}\right\rangle = |\Delta M_{\mathrm{acc}}|/|\Delta M_{\mathrm{don}}| \approx 0.75$. This is due to the short orbital period of the binary, as tidal interactions allow mass transfer to remain fairly conservative. Due to the increase in mass, the accretor now has a luminosity of $\mathrm{log_{10}}(\mathrm{L}/\Ldot) \approx 5.6$ and a surface temperature of $39400\,\mathrm{K}$ (the end of the highlighted red phase in the bottom panel of Fig. \ref{fig:xrb_fiducial_hr}).

The orbital period of the binary only widened from an initial period of $3\,\mathrm{days}$ to $\sim 3.7\,\mathrm{days}$ post-MT. The donor star, which still has a thin hydrogen envelope of $\sim 4.7\,\Mdot$, rapidly expands on a thermal timescale during hydrogen shell burning \citep[e.g,][]{2017A&A...608A..11G, 2020A&A...637A...6L}, and again overflows its Roche Lobe. As this occurs after the initial Case A mass transfer, this phase is referred to as Case AB (with the B referencing mass transfer occuring after the MS but before helium depletion \citealp{1967ZA.....65..251K}). This phase is highlighted in green in Fig. \ref{fig:xrb_fiducial_hr}. The donor loses an additional $\sim 3.5\,\Mdot$, bringing its total mass down to $18.4\,\Mdot$. Tides prevent the accretor from reaching critical rotation, and it is again able to accrete most of the transferred mass, growing to $\sim 44\,\Mdot$. This phase ends just after $\approx 10000\, \mathrm{yr}$, which roughly corresponds to the thermal timescale of the donor. It is now even more stripped of hydrogen (surface mass fraction $\sim 0.18$), and starts contracting towards higher surface temperatures.

Around the same time, it ignites helium in its center, and the core-helium burning phase lasts $\approx 0.44\,\mathrm{Myr}$. During this phase, it has a hot helium rich surface, and would resemble a Wolf-Rayet\footnote{Note that while Wolf-Rayet is a spectroscopic class, we do not compute synthetic spectra. In our stellar evolution models, we switch to Wolf-Rayet wind mass loss rates based on the surface temperature and helium abundance, as defined in Section \ref{sec:mesa_xrb-methods}.} star \citep[e.g,][]{2020A&A...634A..79S} with a strong optically thick wind. These mass loss rates are of the order of $10^{-6} - 10^{-5} \,\Mdot/\rm yr$, and the donor star loses an additional $\sim 6\,\Mdot$ during the WR phase until helium depletion. At this point, it is just a few thousand years away from core-collapse \citep[e.g,][]{1978ApJ...225.1021W}, and we terminate its evolution in \textsc{MESA}. It has a mass of $\sim 12.5\,\Mdot$, with a Carbon-Oxygen (CO) core mass of $\sim 10.6\,\Mdot$ and a central $\rm ^{12}C/^{16}O \sim 0.38$. There is no hydrogen left on the surface, which mostly consists of helium, carbon and some oxygen. Meanwhile, the initially less massive star (representing the current visible star in the XRB) has a mass of $43.4\,\Mdot$. The orbital period of the binary has increased to $\sim 6\,\mathrm{days}$, primarily due to the orbital widening caused by wind mass loss from the donor during its core-helium burning phase \citep{2000A&A...360..227N}.

The pre-SN mass of the progenitor of the compact object, and the pre-SN orbital period align with our expectations of its pre-SN properties from the Monte Carlo simulations in Section \ref{sec:monte-carlo-results}. The mass of the accretor in our models, which corresponds to the visible star in the HMXB is also consistent with its mass estimated through observations.

At the end of the donor star's life, it may successfully explode in a stripped envelope SNe and leave behind a $2 \, \Mdot$ compact object (see Section \ref{sec:explodability} for a discussion on explodability). Once the accretor grows large enough that it is close to filling its Roche Lobe, the system would resemble the HMXB 4U 1700-37 that we observe today. Due to previous mass accetion, we find that the surface of the companion star would appear to be enriched in helium and CNO-processed nitrogen, with surface mass fractions of $\approx 0.42$ and $\approx 4\times 10^{-3}, $respectively. In Section \ref{sec:discussion}, we discuss certain limitations of our fiducial model, and provide further justification for why this system likely formed through Case A MT.

\section{Can the HMXB 4U 1700-37 form a GW source in the future?} \label{xrb-future}

Once the companion star HD 153919 in the HMXB fills its Roche Lobe\footnote{The X-rays observed today are likely from wind accretion, as expected in most HMXBs.}, it will start transferring mass onto the compact object. However, given the highly asymmetric mass-ratio of the system ($q \approx 1/20$), this is likely to become unstable (see also Section \ref{stable}, \citealp{2019A&A...628A..19Q, 2025arXiv251110728O}). We find that the mass transfer rate exceeds $0.1 \, \Mdot/\mathrm{yr}$, which generally signifies the onset of dynamical instability \citep{2025arXiv251009745K}. We assume such systems to undergo a Common Envelope (CE) phase \citep{2023ApJS..264...45F}.

We do not attempt to model the CE phase directly with \textsc{MESA}, and instead utilize the $\alpha \lambda$ formalism \citep{webbink:84,1990ApJ...358..189D} to study its outcome. This depends on the binding energy of the envelope of the star filling its Roche Lobe, as well as energy sources such as the orbital energy that can be used to unbind the envelope. The effective energy of the envelope, $E_{\text{env}}$, depends on the density structure and internal energy of the star, and is given by--

\begin{equation}\label{eq:BE_def}
    E_{\text{env}}(m) = \int_m^M \text{d}m' \left(-\frac{\text{G}m'}{r(m')} + \alpha_{\text{th}}u(m')\right)
\end{equation}

Here, $m$ refers to the mass coordinate corresponding to the
core-envelope boundary, and $M$ is the total mass of the star. The
first term in the integral refers to the gravitational potential
energy of the mass shell at coordinate $m'$. In the second term, $u$ refers to the available internal energy in that shell, and $\alpha_{\text{th}}$ is the fraction of internal energy that can be used to unbind the envelope. If such sources are available to eject the envelope, it reduces the gravitational binding energy and we refer to their sum as the effective energy of the envelope.

The core-envelope boundary is not well defined, and is sensitive to overshooting and other mixing processes in single stars \citep[e.g,][]{2020cee..book.....I}. In binaries, rejuvenation plays an important role in determining the boundary location \citep{2023ApJ...942L..32R,2025ApJ...979...57L}. To account for these uncertainties, we use three different definitions for determining its location, following \cite{2023ApJS..264...45F}. We define the core-envelope boundary as the outermost mass coordinate ($m = M_{\text{core}}$) where the hydrogen mass fraction $X$ is less than $0.01,0.1,0.3$, respectively. The effective envelope energy is often re-parameterized
using $\lambda$ \citep{1990ApJ...358..189D}, defined as--

\begin{equation}
    E_{\text{env}} = -\frac{GM_*M_{\text{env}}}{\lambda R_{*}}
\end{equation}

where $M_*$ is the mass of the star, $M_{\text{env}} \equiv M_* - M_{\text{core}}$ is the mass of its envelope, and $E_{\text{env}} \equiv E_{\text{env}}(m = M_{\text{core}})$.

We can then quantify the outcome of a CE phase using energy conservation, given by

\begin{equation}
    \alpha_{\mathrm{CE}}\,\Delta E_{\mathrm{orb}} = E_{\mathrm{bind}}
    \label{eq:ce_energy_balance}
\end{equation}

\begin{equation}
    \alpha_{\text{CE}}\left(\frac{G\,M_{\mathrm{core}}\,M_{\mathrm{CO}}}{2a_f}
    - \frac{G\,M_*\,M_{\mathrm{CO}}}{2a_i}\right) = \frac{G\,M_*\,M_{\mathrm{env}}}{\lambda\,R_*}.
    \label{eq:ce_energy_terms}
\end{equation}

where $\alpha_{\text{CE}}$ refers to the fraction of orbital energy that can be used to unbind the envelope, $M_{\mathrm{CO}}$ refers to the mass of the compact object, $a_i$ and $a_f$ refer to the orbital separations before and after the CE phase, respectively, and $E_{\mathrm{bind}} = |E_{\mathrm{env}}|$ refers to the effective binding energy of the envelope.

We consider a CE phase to be successful if the post-CE orbital separation $a_f$ is large enough such that the remnant core of the star that loses its envelope $R_{\text{core}} \equiv R(m = M_{\text{core}})$ fits inside its Roche-Lobe. This can be rephrased as the minimum $\alpha_{\text{CE}}$ needed for the remnant core to just fill its Roche-Lobe post-CE. In our fiducial model, if the required $\alpha_{\text{CE}}$ is less than 1, we define the CE phase to be successful, while for values greater than 1, we assume that there is a premature merger between the compact object and the companion's core before the envelope can be ejected. However, $\alpha_{\text{CE}} > 1$ does not necessarily imply a failed CE phase, as there may be additional energy sources such as energy released by accretion onto the compact object (see \citealp{2020cee..book.....I} for a review) that may help eject the envelope. We neglect any potential response of the remnant core to the CE phase, that is $R_{\text{core}}$ does not change, and energy losses via radiation.

We save the internal structure of the
companion star in our fiducial model for the HMXB 4U 1700-37/ HD 153919 at the onset of RLOF, and compute the binding energy of
its envelope. The star is still on the MS and has a central hydrogen mass fraction of $\approx 0.22$.
Therefore, using the $X=0.3$ definition, we
get the core-envelope boundary to be located at $\approx
28.6\,\Mdot$, which roughly lies at the boundary of the present day  convective MS
core of the star. Using this boundary location, and the most
optimistic assumption that all of the internal energy in the envelope
is available, we get the effective binding energy of the envelope to
be $\approx 2.35\times 10^{50}\,\mathrm{erg}$. This corresponds to a
$\lambda \approx 0.5$. Using the energy balance Eqn.
\ref{eq:ce_energy_balance}, we find that the minimum
$\alpha_{\mathrm{CE}}$ that would be required for a successful CE phase is
$\approx 18$. Such an $\alpha_{\mathrm{CE}}$ is quite large, and implies that the CE
phase is likely going to fail to eject the envelope, and lead to a
premature merger between the core of the star and the compact object, in agreement with previous work \citep[e.g,][]{1994MmSAI..65..359P, 2017MNRAS.471.4256V, 2017ApJ...846..170T, 2020A&A...634A..49H}. Although
such systems are not GW sources for ground-based detectors, they could appear as energetic transients (see \cite{2025Ap&SS.370...11G} for a review on other possible multi-messenger signatures) such as LFBOTs \citep{2025arXiv251009745K} or form a Thorne-\.Zytkow Object \citep[e.g,][]{thorne:75, 2025A&A...694A..83N,TZO_review}.

\section{Formation of GW190814} \label{sec:gw190814}

In the previous section, we constructed detailed binary evolution
models at solar metallicity to demonstrate the past evolutionary history of the most asymmetric HMXB in the Galaxy. However, we found that it is unlikely to lead to the formation of a highly asymmetric GW source, as the system will prematurely merge in a future CE phase due to the star's high envelope binding energy. In this section, we compute another set of detailed binary evolution models, and demonstrate scenarios in which a CE phase is successful, and leads to the formation of a highly asymmetric GW source such as GW190814.

The progenitor binaries of the GW events detected by the LVK formed when the universe was much younger, and therefore likely metal poor. \cite{2025ApJ...979..209V} find that BNS formation is not metallicity dependent, while BBH formation is more efficient at lower metallicity.
For these reasons, to describe the formation of a GW190814-like binary that consists of a massive BH,
we compute \textsc{MESA} models at a metallicty of $Z = Z_\odot/10 = 0.00146$.

A key difference in stellar evolution at lower metallicity is that the
mass lost through line-driven winds becomes signifcantly weaker
\citep[e.g.,][]{2001A&A...369..574V,2007A&A...473..603M}. In the
Section \ref{sec:mesa-xrb-results}, we found that the progenitor of the compact object lost a significant amount of mass during its evolution, particularly during its core-helium burning phase. This was crucial in lowering its pre-SN mass down to what we inferred from our Monte Carlo simulations, and potentially aid in its explodability.

Since we now compute binary evolution models at a lower metallicity, we also lower the initial masses of our stars. However, we emphasize that the broad evolutionary pathway remains the same. Based on an initial exploratory work, our fiducial GW190814-like model consists of a binary with initial parameters $27\,\Mdot+25\,\Mdot$ in a 3.5 day orbit. Donors that are substantially more massive than $27\,\Mdot$ will result in pre-SN masses that are potentially too large to successfully explode. On the other hand, accretors that are much less massive than $25\,\Mdot$ will leave behind smaller BHs, which will not appear as GW sources that are \textit{highly asymmetric} in their mass-ratio.

\subsection{Evolution before first-SN}

\begin{figure}[htbp]
    \centering
    \includegraphics[width=0.47\textwidth]{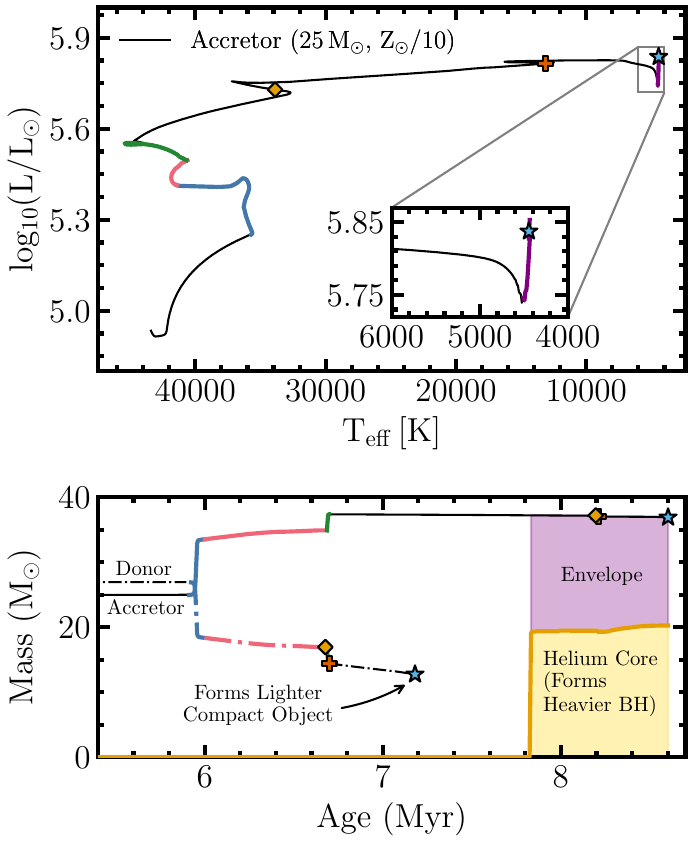}
    \caption{The evolution of the accretor in our fiducial model for forming a GW190814-like binary at a $Z = Z_\odot/10 = 0.00146$. The initial masses of the donor and accretor are $27\,\Mdot$ and $25\,\Mdot$, and the initial orbital period is $3.5\,\mathrm{days}$. \textit{Top panel:} Evolution of the accretor on the HR diagram. The colors and marker symbols follow the same style as in Fig. \ref{fig:xrb_fiducial_hr}. The inset panel zooms into the late stages of the core-helium burning phase, where it ascends the RSG branch, and developes a convective envelope (shown in purple). \textit{Bottom panel:} The evolution of the mass of the donor (dash-dot) and accretor (solid) over time. The former forms the lighter compact object, while the latter has a massive helium core (shown in golden), and is the progenitor of the massive BH in an asymmetric mass-ratio GW190814-like event.}
    \label{fig:gw_hr_accretor}
\end{figure}

In Fig. \ref{fig:gw_hr_accretor}, we show the evolutionary track of the accretor in the binary on the HR diagram (top panel), as well as the evolution of their masses over time (bottom panel).

As stars at lower metallicity are more compact, mass transfer begins later in the evolution. After $5.9 \,\mathrm{Myr}$, the donor star overflows its Roche-Lobe during the MS, initiating Case A mass transfer. This phase lasts for around $0.75 \,\mathrm{Myr}$. Once the donor reaches TAMS, it starts contracting, and detaches itself from its Roche-Lobe. At this stage, the donor star has a mass of $16.9\,\Mdot$, while the accretor has grown to about $34.9\,\Mdot$. Due to the short orbital period, strong tidal forces prevent the accretor from reaching critical rotation, and mass transfer remains nearly fully conservative. The donor soon re-expands once it starts hydrogen shell burning, initiating the second phase of Case AB mass transfer, similar to our fiducial model in Section \ref{sec:mesa-xrb-results}. At the end of this phase, the binary consists of a $14.4\,\Mdot$ donor, and a $37.4\,\Mdot$ accretor in a tight $6.4\,\mathrm{day}$ orbit. Around the same time, the donor contracts (ending MT) and ignites helium in its core, which lasts for $\sim 0.5 \, \rm Myr$. At core-helium depletion, the donor is left with a mass of $12.7\,\Mdot$, with a carbon-oxygen core mass of $10.7\,\Mdot$ (and a central $\rm ^{12}C/^{16}O \sim 0.39$). On the other hand, the accretor, now the more massive star in this system has a mass of $37.3\,\Mdot$.

Since we terminate our donor models at core helium depletion, we neglect the impact of binary and tidal interactions after this point. We assume that at least in some cases, the donor star successfully explodes at
the end of its life, and leaves behind a NS, or a low mass BH due to
fallback or accretion during the SN. This may resemble the
lower mass compact object in GW190814. We discuss the effects of explodability in more detail in Sec.~\ref{sec:explodability}. The compact object may also
receive a natal kick at birth, in analogy with the case of the HMXB 4U 1700-37. Depending on its strength and direction, a variety of post-SN outcomes are possible. This is unlike the Galactic HMXB 4U 1700-37, which was bounded by observational constraints on the post-SN orbit.

\subsection{Evolution after first-SN}

For systems that remain bound post-SN, the binary consists of a
compact object and a companion star, which was originally the
accretor. Depending on the size of the orbit post-SN and the radial
evolution of the companion star, it may (or may not) fill its Roche-Lobe during its evolution.

If it does fill its Roche-Lobe, a phase of reverse-RLOF onto the compact object will ensue. The outcome of this phase sensitively depends on the evolutionary state of the companion star and the stability of mass transfer at the onset of RLOF. The highly asymmetric mass-ratio of the system ($q = 2.6/37.3 \approx 1/14$) is similar to the mass-ratio of the Galactic HMXB that we described in Section \ref{sec:monte-carlo-results}. We again assume that mass transfer becomes unstable and leads to a CE phase.

\subsubsection{Fate of CE phase}

We use the $\alpha \lambda$ formalism, as described in Section
\ref{xrb-future} to determine the outcome of the CE phase. However, as
the post-SN orbital separation can take a range of values, the density
structure of the star, and correspondingly the binding energy of its
envelope at the onset of CE could vary significantly.

To account for the different possible outcomes post-SN, we ``detach" the companion star from the compact object, and follow its evolution as a single star in \textsc{MESA} (following \citealp{2021ApJ...923..277R,2023ApJ...942L..32R}). This allows us to model the evolution of its density structure and envelope binding energy along with its radial evolution, thereby accounting for different possible post-SN orbital separations\footnote{We neglect the impact of tides and potential irradiation from an accreting compact object in orbit around the star.}.

\begin{figure}[htbp]
    \centering
    \includegraphics[width=0.47\textwidth]{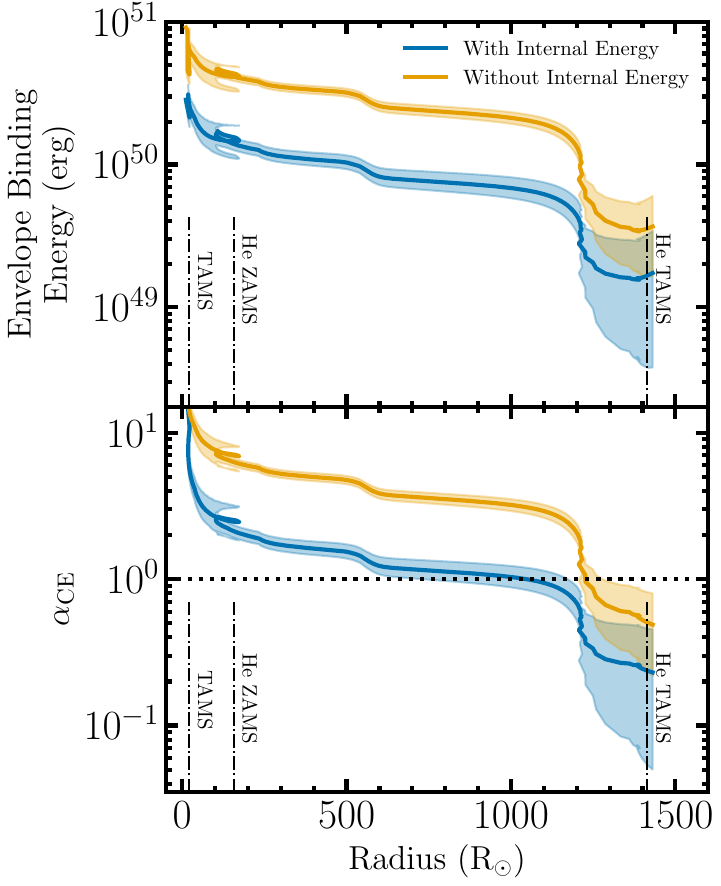}
    \caption{The need for stellar radii $\gtrsim 1250\,\Rdot$ for a successful Common Envelope phase ($\alpha_{\text{CE}} < 1$). \textit{Top panel:} The evolution of the binding energy of the envelope as a function of the total radius of the star. Blue curves include the contribution of internal energy, while golden curves only include the gravitational binding energy. The solid lines correspond to the $X = 0.1$ definition for the core-envelope boundary, while the shaded bands encompass the $X = 0.01$ and $X = 0.3$ limits. The vertical dash-dot lines denote the stellar radius when it reaches TAMS, ignites helium in its core (He ZAMS), and finishes core-helium burning (He TAMS), respectively. \textit{Bottom panel:} The minimum $\alpha_{\text{CE}}$ required to eject the envelope as a function of the stellar radius, with the same color and shading conventions as the top panel.}
    \label{fig:gw_be}
\end{figure}

In the top panel of Fig. \ref{fig:gw_be}, we show the effective
binding energy of the envelope of the companion star as a function of
its total stellar radius, which we use as a proxy for its temporal
evolution.
For most of its evolution, as the star grows in radius, there is a moderate decrease in its binding energy. Notably, our models ignite helium in their core while crossing the Hertzprung-Gap, and spend most of the core-helium burning phase in a blue loop (see Fig. \ref{fig:gw_hr_accretor} and Fig. \ref{fig:gw_be}, \citealp{2023ApJ...942L..32R}). However, during the end of its core-helium burning phase, when it has grown to a size of almost $1250\,\Rdot$, it developes a loosely bound convective envelope and starts ascending the Red Super Giant (RSG) branch (highlighted in purple in Fig. \ref{fig:gw_hr_accretor}). This dramatically decreases the effective binding energy of its envelope, making it easier to eject during a potential CE phase \citep[e.g,][]{2021A&A...645A..54K}.

Using the effective binding energy of the star's envelope, and the
method described in Section \ref{xrb-future}, we calculate the minimum
$\alpha_{\text{CE}}$ that would be required to eject it as a function of its stellar radius. This is shown in the bottom panel of Fig. \ref{fig:gw_be}, where the line colors and shaded bands follow the same convention as the top panel. For most of its evolution, when $R_* < 1250\,\Rdot$, the minimum $\alpha_{\text{CE}}$ required to successfully eject the envelope is much greater than one. These correspond to the cases similar to the HMXB 4U 1700-37, where we assume that the CE phase would fail, and there would be a premature merger between the compact object and the core of the star.

However, for $R_* > 1250\,\Rdot$, we find that the minimum $\alpha_{\text{CE}}$ required to eject falls below one. In such scenarios, we assume that the CE phase is successful in ejecting the envelope, and would leave behind a tight, and potentially circularized binary consisting of the remnant core of the star that lost its envelope, and the compact object.

\subsubsection{Forming a highly asymmetric mass-ratio double compact system}

In our \textsc{MESA} models, we find that for the systems that survive the CE phase, their remnant cores just have $\approx 0.03\,\mathrm{Myr}$ left to the end of core-helium burning phase. They have a helium core mass of $\approx 20\,\Mdot$, and radius $\approx 0.8\,\Rdot$, which would appear as a WR star. For an $\alpha_{\mathrm{CE}} = 1$, the post-CE orbital separation is $\approx 2-5\,\Rdot$, depending on whether we exclude or include internal energy. In either case, the remnant core comfortably fits inside the new Roche-Lobe. Due to the short time left, we assume that the mass of the core will not change significantly. It may appear as an X-ray source, similar to Cygnus X-3, which consists of a massive WR star in a tight orbit with a compact companion \citep[e.g,][]{2013ApJ...764...96B,2024NatAs...8.1031V}. At core-collapse, such a heavy core in our fiducial model is potentially too large to explode. It will likely implode on itself with little to no mass loss, leaving behind a $\approx 20\,\Mdot$ BH. Assuming that such massive BH's receive negligible natal kicks, these systems would now consists of a NS or a low-mass BH with a massive $\approx 20\,\Mdot$ BH companion in a tight circular orbit of a few solar radii. At this time, the compact binary emits GWs at a frequency of $\approx 0.1\,\mathrm{mHz}$, which lies in the LISA frequency range \citep{2023LRR....26....2A}, and may be a detectable source. The subsequent evolution of this system will be dominated by a GW-driven inspiral. For post-CE orbital separations of $2-5\,\Rdot$, we find that it takes at most $\approx 80\,\mathrm{Myr}$ for the inspiral to end in a finale as a compact binary coalescence that may be observed by the LVK detectors \citep{Peters:1963ux}. These signals would consist of a low mass compact object that may be a NS or BH, with a massive BH companion, that would appear similar to the highly asymmetric mass-ratio event GW190814. Since these are post-CE systems and the formation of the massive BH likely did not induce any significant eccentricity in the binary, we do not expect any measureable eccentricity at merger. Additionally, we also do not expect the second born massive BH to have any significant spin, which is also consistent with the estimated spin of the primary BH in GW190814. We discuss this in more detail in Section \ref{spin}.

 \section{Rates of GW190814-like events} \label{rates}

 In Section \ref{sec:mesa-xrb-results}, we demonstrated that the highly asymmetric mass-ratio Galactic HMXB 4U 1700-37 can form through isolated binary evolution, starting from a relatively equal mass-ratio binary that undergoes conservative mass transfer. In Section \ref{sec:gw190814}, we showed that a similar formation pathway works at lower metallicity as well, and leads to the formation of a highly asymmetric mass-ratio GW190814 like source if the CE phase is successful. Assuming an evolutionary link between the two systems, we can utilize the number of highly asymmetric mass-ratio Galactic HMXBs observed to estimate the rate of highly asymmetric mass-ratio GW events like GW190814. There exist two such HMXBs in the Galaxy. One of them is the Galactic HMXB 4U 1700-37 that we modeled in Section \ref{sec:mesa-xrb-results}. The other is the X-ray pulsar GX 301-2, which has a $\approx 36-50\,\Mdot$ B supergiant companion, Wray 977 \citep{1995A&A...300..446K}. Assuming that the number of Milky Way equivalent galaxies (MWEG) in the local Universe is $\sim 1.16 \times 10^7 \,\rm{Gpc}^{-3}$ \citep{2008ApJ...675.1459K,2010CQGra..27q3001A}, and that the average lifetime of observing a binary in the HMXB phase is $\approx 0.1\, \rm Myr$ \citep{2023pbse.book.....T}, the local rate of observing highly asymmetric mass-ratio HMXBs is given by--

 \begin{align}
    \rm \mathcal{R}_{\mathrm{HMXB}} &\sim \frac{2}{\mathrm{MWEG}} \times \frac{1.16 \times 10^7\, \mathrm{MWEG}}{\mathrm{Gpc}^{-3}} \times \frac{1}{0.1\, \mathrm{Myr}} \\
    &= 232\, \mathrm{Gpc}^{-3} \mathrm{yr}^{-1}
 \end{align}

\begin{figure}[htbp]
    \centering
    \includegraphics[width=0.47\textwidth]{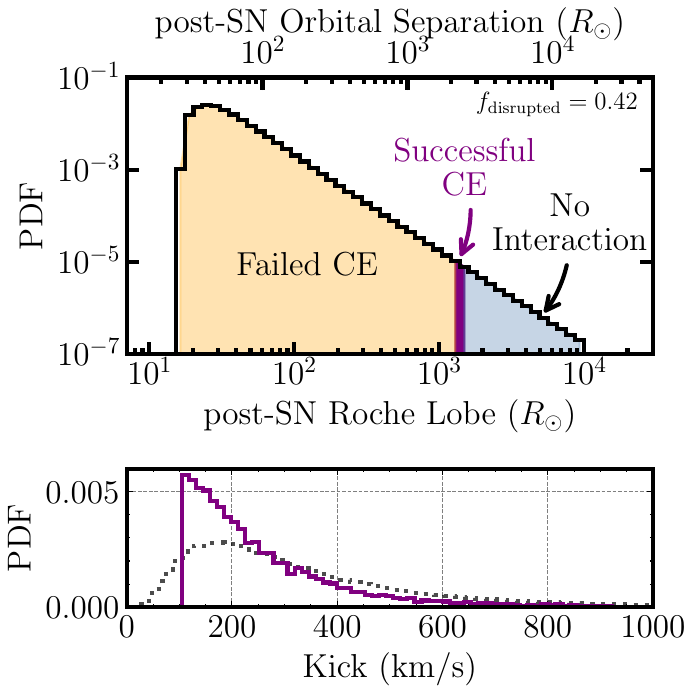}
    \caption{Only $\approx 0.24 \%$ of asymmetric mass-ratio HMXBs form asymmetric mass-ratio GW sources, all of which need natal kicks $\gtrsim 100\,\mathrm{km/s}$ to the first-born compact object. \textit{Top panel:} The distribution of post-SN Roche Lobe sizes for our
fiducial model for the formation of a GW190814-like system, assuming the natal kicks to follow a lognormal distribution. Yellow band corresponds to
post-SN Roche Lobes smaller than $1250\,\Rdot$, which may appear as highly asymmetric Galactic HMXBs, but are unlikely to form a GW
source in the future due to a failed CE phase. Purple band corresponds to post-SN
Roche Lobes between $\approx 1250-1500\, \Rdot$, which
have a successful CE phase and form a highly asymmetric mass-ratio GW190814-like source. Blue band denotes orbits large enough such that the star does not fill its Roche Lobe during its subsequent evolution. The top axis denotes the corresponding orbital separation, for the assumed mass-ratio of $q = 2.6\,\Mdot/37.3\,\Mdot \approx 1/14$. \textit{Bottom panel:} The distribution of natal kicks for a successful CE phase (in purple), while the assumed lognormal prior is shown in dotted black.}
    \label{fig:gw_post_SN_rl}
\end{figure}

 However, as we demonstrated in Sections \ref{sec:mesa-xrb-results}
and \ref{sec:gw190814}, not all highly asymmetric mass-ratio HMXBs
will form GW190814-like events. We found that a successful CE phase
occurs if the companion star fills its Roche-Lobe when it is a RSG
with a loosely bound convective envelope. In our fiducial model, this requires the size of
the Roche Lobe after the first-SN to be $\approx 1250-1500\, \Rdot$.
In Fig. \ref{fig:gw_post_SN_rl}, we show the distribution of possible
Roche-Lobe sizes after the first-SN (calculated using fits from \citealp{1983ApJ...268..368E}). Based on our fiducial model, we assume that a $12.7\,\Mdot$
star explodes to leave behind a $2.6\,\Mdot$ compact object, and it
has a companion of $\approx 37\,\Mdot$ in a $\approx 7\,\mathrm{day}$
orbit. The only free parameters that determine the size of
the post-SN RL are the magnitude and direction of the natal kick. Similar to Section \ref{sec:monte-carlo-method}, we
assume isotropic natal kicks ($v_{\mathrm{kick}}$) to arise from a
lognormal distribution \citep{2025ApJ...989L...8D}. We find that $\approx 42\%$ of systems are disrupted due to the SN. For the systems that remain bound, the yellow shaded band corresponds to post-SN Roche Lobes that are smaller than $1250\,\Rdot$, which would only appear as a highly asymmetric mass-ratio HMXB, but not form a GW source due to a failed CE phase. The purple shaded band corresponds to post-SN Roche Lobes that are $\approx 1250-1500\, \Rdot$, which will have a successful CE phase and form a highly asymmetric mass-ratio GW source. The blue shaded band corresponds to orbits that are so large that the star never fills its Roche Lobe. For the assumed kick distribution, we find that only $\approx 0.24\%$ of systems that appear as a HMXB can form a GW source through a successful CE phase. Notably, the first-born compact objects in those systems all receive natal kicks greater than $100\,\mathrm{km/s}$ (bottom panel of Fig. \ref{fig:gw_post_SN_rl}), similar to what we needed to explain the HMXB 4U 1700-37 in Section \ref{sec:monte-carlo-results}. Thus we can estimate the local rate of highly asymmetric mass-ratio GW190814-like events to be

 \begin{equation}
    \rm \mathcal{R}_{\mathrm{GW}} \sim 0.0024 \times \mathcal{R}_{\mathrm{HMXB}} \sim 0.56\, Gpc^{-3} yr^{-1}
 \end{equation}

 This estimate is within an order of magnitude of the
 $\rm 1-23\, Gpc^{-3} yr^{-1}$ rate of GW190814-like events estimated
 by the LVK collaboration \citep{2020ApJ...896L..44A}, which is based on a single event (GW190814). However, we caution that our rate estimate depends on key uncertain factors such as the lifetime of the HMXB phase, and the radial extent of the star for a successful CE phase. A detailed rate calculation also requires the use of
 population-synthesis techniques to account for the star formation
 history, metallicity dependence, and other uncertainties in linking
 Galactic HMXBs to GW events \cite[e.g.,][]{2020ApJ...899L...1Z, 2023ApJ...953..152O, 2025arXiv251116648M}. The observed rate of highly asymmetric mass-ratio GW events may also include contributions from other formation scenarios, both within the context of isolated binary evolution, as well as other pathways such as dynamical interactions or triple evolution.

\section{Discussion} \label{sec:discussion}

In this section we discuss additional observational constraints for the HMXB 4U 1700-37/ HD 153919, and certain assumptions and caveats for our corresponding models. We discuss the explodability of the first-born compact object, the stability of the second phase of mass transfer and the expected spins of the massive BH in our models for GW190814. We also compare our results to previous studies that explored the origin of GW190814-like events.

\subsection{Kinematic Age of the HMXB 4U 1700-37} \label{age}

Using kinematic and distance measurements with \emph{Gaia},
\cite{2021A&A...655A..31V} estimated that the HMXB 4U 1700-37 has a
kinematic age of $\approx 2.2\,\mathrm{Myr}$. This information can also be used to constrain its evolutionary history. Our models for the companion star HD 153919 agree with its estimated mass and observed surface properties, such as its effective temperature and luminosity. However, we find a discrepancy in the time it takes to evolve to its current state post-SN (i.e, its kinematic age). In our models, the companion star has  $\approx 1\,\mathrm{Myr}$ left on the MS after the SN, and only $\approx 0.7\,\mathrm{Myr}$ to reach the observed position of HD 153919 on the HR diagram. This implies a factor of $2-3\times$ discrepancy in the kinematic age of the system. Although unresolved, we list a few potential explanations below--

\begin{enumerate}
    \item \textit{Higher systemic velocity in the past:} This would imply a smaller kinematic age. This could happen if the system has slowed down over time as it escapes the cluster potential well.
    \item \textit{Larger initial donor mass:} This would imply a shorter lifetime for the progenitor of the compact object, thereby extending the time spent by the companion star in the post-SN phase. However, a larger donor mass would also imply a slightly larger pre-SN mass of the progenitor of the compact object, which may be inconsistent with what we inferred from the Monte Carlo simulations in Section \ref{fig:xrb_monte_carlo}.
    \item \textit{Smaller initial accretor mass:} This would imply a smaller mass for the star post-SN, increasing its MS lifetime (see Appendix \ref{grid}). However, this could be in tension with the estimated luminosity and mass of the companion star, as we find that (initially) lower mass accretors are less luminous and less massive at present day than what is inferred from observations.
    \item \textit{Stronger core rejuvenation:} As the accretor gains mass, its core gets rejuvenated, which increases its MS lifetime. If rejuvenation is stronger than implemented in our evolutionary models, it may increase its kinematic age.
\end{enumerate}

The discrepancy could also be addressed by a combination of the above factors. Previous studies have also found a 2$\times$ discrepancy in the ages as measured from single star evolutionary grids and kinematic ages \citep[e.g.,][]{2018A&A...619A..78L}, and further studies are needed to address this issue.

\subsection{Rotational Velocity of HD 153919} \label{vrot}

HD 153919 is the visible companion star to the X-ray source 4U 1700-37. As it is likely a product of past mass transfer, it may have a high surface rotational velocity. However, \cite{2020A&A...634A..49H} used spectral observations and modeling to estimate an upper limit on its projected rotational velocity of $110^{+30}_{-50}\,\mathrm{km/s}$. This implies that HD 153919 is currently sub-synchronous with the $3.41\,\mathrm{day}$ orbit.
However, in our models, tidal forces are strong enough to keep the companion star sychronized to the orbital period, and as a result have a large surface rotational velocity of $\approx 300-400\,\mathrm{km/s}$. However, various studies \citep[e.g,][]{2018A&A...616A..28Q, 2023ApJ...952...53M} have suggested that tidal effects may not be as large as currently implemented (following \citealp{1977A&A....57..383Z} as implemented in \citealp{2002MNRAS.329..897H}). By turning off tidal forces in our models, we find much better agreement with the observed surface rotational velocity of the companion star. In the first phase of mass transfer prior to SN, tidal effects were crucial to keep the accretor star from reaching critical rotation, allowing for nearly conservative mass transfer. The truth likely lies somewhere in between. The exact strength of tidal forces is currently uncertain, and depends sensitively on the internal structure of the star. The surface rotational velocity of HD 153919 depends on a complex interplay of AM transport during MT, the effects of time varying tidal effects, and angular momentum lost through winds that are not captured in full detail by our models, and warrants further study, along with more observational constraints.

\subsection{Does the HMXB 4U 1700-37 need to form through Case A mass transfer?} \label{caseA}

\textit{Common Envelope:} Given the short period of the HMXB today, it could also be a post-CE system. However, the companion star HD 153919 has an estimated mass of $\approx 40-50\,\Mdot$. Since it likely did not gain appreciable mass during a CE phase, the progenitor of the compact object must have been much more massive than $\approx 40-50\,\Mdot$, as unstable mass transfer generally occurs in systems with an asymmetric mass-ratio. For a CE phase to be successful, the progenitor of the compact object must have been in the post-MS phase with a well developed core-envelope boundary. Thus, given its high initial mass, its core would potentially be too massive to successfully explode and leave behind a $\approx 2\,\Mdot$ compact object as is seen in the system today. Additionally, based on the results of our Monte Carlo simulations described in Section \ref{sec:monte-carlo-results}, we know that the pre-SN mass of the progenitor of the compact object was less than $13\,\Mdot$, which is far too small of a core for a star that had an initial mass more than $\approx 40-50\,\Mdot$. Thus, a CE phase in the past is unlikely to explain the observed properties of the HMXB.

\textit{Case B:} Another possibility is that the first phase of MT remained stable, but
occured after the MS of the donor, that is Case B.
However, this is unlikely to explain the short period of the system today. Based on our Monte Carlo simulations, the orbital period pre-SN must have been $<10\,\mathrm{days}$. For the mass range relevant to the system of study, Case B occurs for initial periods that are much longer than $10\,\mathrm{days}$ \citep[e.g,][]{2023A&A...672A.198S}. Additionally, \cite{2019A&A...624A..66R} showed that for Case B mass transfer, the orbital period of the binary post-stable MT is longer than its initial period in most cases, except for scenarios where mass transfer is non conservative ($\beta < 0.6$) \textit{and} there are extreme losses in angular momentum ($\gamma_{\rm RLOF} \geq 2$). Therefore, we find that it is exceedingly difficult for a binary post-Case B MT to shrink significantly post-SN and explain the system's short orbital period today. Additionally, even for Case B, the progenitor of the compact object would already have a well developed core before MT occurs, and would be potentially too massive to successfully explode and form a $\approx 2\,\Mdot$ compact object.

\subsection{Explodability of first-born compact object}\label{sec:explodability}

In Sections \ref{sec:mesa-xrb-results} and \ref{sec:gw190814}, we
described an evolutionary pathway to forming highly
asymmetric mass-ratio Galactic HMXBs and GW sources. In both scenarios, we require that the donor star in the binary successfully explodes at the end of its life, and leaves behind a
low mass compact object which could either be a NS or a low mass BH. Observations of the Galactic HMXB 4U 1700-37 strongly suggest that this pathway can occur in nature, as this system, which originated in a very young cluster, consists of a low mass compact object with a massive stellar companion.

However, the explodability of massive stars is sensitive to their pre-SN structure and past evolutionary history such as mass
lost through winds or binary interactions. Various 3D simulations of core-collapse SNe are
converging on the imporance of neutrino-driven convection in
 driving successful explosions \citep[e.g,][]{2021Natur.589...29B, 2025arXiv250214836J}. However, they are
computationally intensive and can only be done for a few progenitor models at a time, which are generally computed with 1D stellar evolution models. At present, there is a lack of large grids of 1D models at core-collapse, that span a wide range of single and binary star evolutionary histories, and accurately compute through the late stages of nucelar burning that require large nuclear networks \citep[e.g,][]{2016ApJS..227...22F, renzo2024progenitorsmallreactionnetworks,2025ApJS..279...49G}. This has led to the development of various simplified prescriptions and explodability criterion \citep[e.g.,][]{Fryer_2012,Patton_2020,2025A&A...700A..20M} that are used in population-synthesis studies to map stars to their remnant properties, such as the mass of the compact object, which can be tied to the natal kick that they recieve \citep[e.g,][]{2020MNRAS.499.3214M}.

In our detailed binary evolution models computed with MESA, we only evolve the stars till core helium depletion. Therefore, we have access to the mass and composition of the helium core, the carbon-oxygen core, and its density profile at that time, which may already determine its explodability \citep[e.g,][]{2025A&A...695A..71L}. Our models have core masses that lie between $\sim 10-13 \, \Mdot$. The explodability of stars with carbon-oxygen core masses between $\sim 7-15 \, \Mdot$ is uncertain, and potentially stochastic \citep[e.g,][]{2025A&A...700A..20M}. If such stars do explode after losing their envelopes, they would appear as stripped envelope SNe, with ejecta masses of $\approx 8-10\,\Mdot$. This is on the higher end of estimated ejecta masses of observed stripped envelope SNe \citep{2016MNRAS.457..328L}. However, recent studies \citep[e.g,][]{2021A&A...645A...5S,2021A&A...656A..58L,2021ApJ...916L...5V,2023ApJ...948..111F} find that donors in binary system may be easier to explode than their single star counterparts \citep[but see also][]{2024RNAAS...8..302K}. Additionally, such stripped envelope SN would have a massive companion star next to it, which may be observable with follow up deep imaging campaigns to search for surviving companions in such supernovae \citep[e.g,][]{2017ApJ...842..125Z, 2026MNRAS.546f2208Z, 2026MNRAS.545f2163S}. On the higher mass end, it is expected that helium cores $> 15\, \Mdot$ are potentially too large to lead to successful explosions (e.g,\citealp{Fryer_2012,Patton_2020,2025A&A...700A..20M}, but see also \citealp{2018MNRAS.477L..80K,2023ApJ...957...68B,2024ApJ...964L..16B,2025ApJ...987..164B})\footnote{With rotation and magnetic fields, $\gtrsim 100\,\Mdot$ carbon-oxygen cores can potentially explode as well \citep{2022ApJ...941..100S,2025arXiv250815887G}.}. Further work is needed in 1D simulations to compute large grids of pre-SN models with large nuclear networks that span a wide range of single and binary evolutionary histories, as well as a large grid of 3D simulations to better understand the landscape of explodability and develop more accurate prescriptions that can be used in population synthesis simulations. To explicitly confirm whether the first-born compact object in our models for highly asymmetric mass-ratio HMXBs and GW events can be a NS or a low mass BH, we need to evolve our models beyond core helium depletion till core-collapse, which we defer to future work.

\subsection{Spin of the massive BH} \label{spin}

In addition to the masses of the compact objects and the distance to
the source, the GW signal also encodes information on their spins. While it is difficult to measure individual spins,
the GW detectors are sensitive to $\chi_{\mathrm{eff}}$, a mass-weighted combination of the two spins projected onto the orbital angular momentum. For highly asymmetric mass-ratio systems, where $m_1$ is much larger than $m_2$, the GW signal provides a strong constraint on the spin of the massive BH. This is true for GW190814, where the spin of the primary BH is tightly constrained to be $\leq 0.07$ \citep{2020ApJ...896L..44A}.

Spins of compact objects are closely tied to the evolutionary history of their progenitors, such as the angular momentum (AM) transport between their cores and envelopes, the mass and angular momentum gained or lost through winds and binary interactions, and also the effects of tides in close binaries. Based on our current understanding of AM transport within stars, it is expected that the first-born BHs are born with negligible spin \citep[e.g,][]{2019ApJ...881L...1F}. However, the progenitor of the second born BH may get spun up before its death through tidal interactions with its companion (e.g, \citealp{2020A&A...635A..97B}, see also \citealp{2025A&A...696A..54S}).

In our evolutionary scenario for GW190814-like systems, we find that the massive BH is the second born compact object. To first order, if one assumes efficient tidal locking, it may be expected to have a large spin, which would be in tension with the measured low spin of the massive BH in GW190814. However, detailed numerical studies of tides in WR+BH binaries \citep{2023ApJ...952...53M} find that spin up through tides is not be as efficient as previously assumed. The strength of tidal forces scales as $q^2$, thus it also gets progressively smaller as the mass-ratio of the binary decreases. Additionally, our models undergo a CE phase towards the end of core-helium burning, with only $\approx 0.03\,\mathrm{Myr}$ left till helium depletion in our fiducial model. Therefore, even if we disregard the damping of tidal forces due to the asymmetric mass-ratio of the system, there is likely not enough time for tides to spin up the star's remannt core prior to the end of its life. Thus, we expect the massive BH to not have any significant spin, and emphasize that the measured negligible spin of the massive BH in GW190814 need not imply that it is the first-born compact object in the binary.

\subsection{Stability of second phase of mass transfer} \label{stable}

In our fiducial model for both the HMXB 4U 1700-37 and GW190814,
we assume that mass transfer from the companion star to the first-born
compact object is unstable, and leads to a CE phase.
This is primarily based on the extreme mass-ratio of the binary
\citep[e.g,][]{2014A&A...563A..83C,2017MNRAS.465.2092P}.
We also find that the mass transfer rates during this phase exceed
$0.1\,\Mdot/\mathrm{yr}$, which has been used as threshold
indicating the onset of dynamical instability \citep[e.g.,][]{2023ApJS..264...45F}. In our default setup, we assume that mass lost from the system carries with it the specific angular momentum of the accretor's center of mass. However, recently \cite{2025arXiv251110728O} suggested that mass transfer in such extreme mass-ratio binaries may remain stable if mass transfer is assumed to be non conservative, and the mass that is lost from the system carries away the specific angular momentum of the donor's center of mass. In such a scenario, the binary widens during mass transfer, allowing it to remain stable. On implementing such a setup in our binary evolution models, we find that although mass transfer remains stable and avoids a CE phase, the orbit of the binary widens to $\approx 80\,\Rdot$. This is large enough such that when the double compact binary forms, it takes much longer to merge through a GW driven inspiral than the age of the Universe. Therefore, we do not expect such a pathway to contribute to the formation of highly asymmetric mass-ratio GW events. However, it may play a role in the future evolution of the HMXB 4U 1700-37/ HD 153919, and further work is needed to determine the stability of mass transfer in such extreme mass-ratio binaries.

Additionally, \cite{2019A&A...628A..19Q} showed that mass-transfer in HMXBs could remain stable on a nuclear timescale if the donor star has a steep H/He gradient beneath its surface. They suggest that the presence of a gradient can stabilize mass transfer upto mass-ratios of $\approx 1/20$. However, for the gradient to be revealed at the surface, the star must have lost a significant amount of mass in the past. To explain this, \cite{2019A&A...628A..19Q} suggest that the HMXB phase may have been preceded by a prior CE phase that strips its envelope, and reveals the H/He gradient. Such stripped stars would have different surface abundances and spectra than predicted by our models where the companion star was an accretor in the past. In such a scenario, it may also be a challenge to explain the presence of the low mass compact object in the HMXB 4U 1700-37, although further observations of this system and HD 153919 in particular will help distinguish between different formation pathways.

\subsection{Comparison to previous work}

Various studies have aimed at understanding the potential formation
pathways for GW190814-like systems \citep[e.g,][]{2020ApJ...899L...1Z, 2021MNRAS.500.1380M, 2021ApJ...908L..38A, 2021MNRAS.500.1817L, 2025arXiv251116648M}. In the context of
isolated binary evolution, such studies have focused on the mass of
the secondary compact object, and/or the mass-ratio of the event. The
majority of these studies have only explored these questions using rapid
binary population synthesis techniques. In particular,
\cite{2020ApJ...899L...1Z} find two potential scenarios for forming
such systems, one where the low-mass compact object
forms first, and another where the larger BH forms first. The
evolutionary scenario that we demonstrate in this work with detailed binary evolution models is qualitatively
similar to the former where the first born compact object requires a large natal kick to widen the orbit post-SN. However, they find that the rates of such systems are a couple of orders of magnitude lower than the LVK estimated rate of GW190814-like events. Motivated by the existence of a low mass compact object in the highly asymmetric mass-ratio Galactic HMXB 4U 1700-37, we showed that the former scenario can occur in nature and demonstrated its feasibility with detailed binary evolution models (althought see \citealp{2025ApJ...989..188X}). The same formation pathway was also found by \cite{2021MNRAS.500.1380M}, but required Hertzprung-Gap donors to survive the CE phase. This may not be possible, given the high envelope binding energies while crossing the HG. Additionally, as discussed in Section \ref{sec:gw190814}, we find that star should be ascending the RSG branch and nearing the end of its core-helium burning phase for a successful CE phase.

\section{Conclusion} \label{sec:conclusion}

The catalog of GW events from compact binary mergers is expected to rapidly grow in the coming years. Understanding where, and how these sources form is a key goal for studies of massive (and binary) star evolution. However, the masses and spins of the compact objects as inferred from the GW data are often insufficient to infer their formation history, which can involve many intermediate stages. This is particularly relevant for the astrophysical interpretation of the (numerous) exceptional GW events that may be outliers to the rest of the GW population, which is bound to increase as the catalog size increases. In such situations, it can be beneficial to understand the origins of the (numerous) earlier stages in the evolution of massive stars in binaries. Such systems abound in the local Universe, such as Algols \citep[e.g,][]{2025arXiv251115347S}, WR+O systems \citep[e.g,][]{1998NewA....3..443V, 2025A&A...695A.117N}, O+BH binaries (e.g, \citealp{2022NatAs...6.1085S}, see \citealp{2026arXiv260203650D} for a potential candidate), HXMBs \citep[e.g,][]{2023A&A...671A.149F, 2026arXiv260208152K}, DNSs \citep[e.g,][]{2018MNRAS.481.4009V}, companions to stripped envelope supernovae \citep[e.g,][]{2017ApJ...842..125Z, 2026MNRAS.546f2208Z, 2026MNRAS.545f2163S} etc. They offer a treasure trove of observational constraints that can be utilized to understand their evolutionary history, which are inaccessible when solely studying GW sources.

We demonstrate this approach for GW190814, which, with a $q \approx 0.1$ remains the most asymmetric mass-ratio compact binary merger observed by the LVK detectors. Although there have been several proposed formation pathways for this event, its origin remains unclear. Leveraging the highly asymmetric mass-ratio Galactic HMXB 4U 1700-37, we find that:

\begin{itemize}
\item Utilizing several observables such as its orbital properties, kinematics, and local environment, we use Monte Carlo simulations to reconstruct the state of the HMXb 4U 1700-37/ HD 153919 pre-SN. We found that the progenitor of the compact object had a pre-SN mass $<13\,\Mdot$, and the binary had a pre-SN orbital period $<10\,\mathrm{days}$. This strongly suggests that the system underwent a phase of mass transfer in the past, during the Main Sequence of the primary (case A).

\item We construct detailed binary evolution models with \textsc{MESA} to explain the evolutionary history of the HMXB, and find that conservative mass transfer on the Main Sequence (Case A), along with a directed natal kick can explain the observed properties of the system today.

\item We demonstrate that that the HMXB is unlikely to form a GW source in the future. It will likely undergo a failed CE phase in the future due to the high binding energy of the companion star's envelope, in agreement with previous work \citep[e.g,][]{1994MmSAI..65..359P, 2017MNRAS.471.4256V, 2017ApJ...846..170T, 2020A&A...634A..49H}.

\item By computing additional models at lower metallicity, we show that a similar evolutionary pathway can also form highly asymmetric mass-ratio GW sources like GW190814. The first phase of mass transfer remains stable and conservative, which inverts the mass-ratio of the binary. For a successful CE phase in the future, we find that the first born compact object must receive a strong natal kick ($\gtrsim 100\,\mathrm{km/s}$) to widen the orbit post-SN. This allows for the CE phase to occur when the companion star is a RSG with a loosely bound convective envelope that is easier to eject.

\item In both scenarios, we find that the first-born compact object is the lower mass compact object, that is a NS or a low mass BH. For the systems that form a GW source, the massive BH is the second born compact object, and we expect it to have negligible spin, in agreement with measured spin of the primary (more massive) BH in GW190814.

\item Utilizing the fact that there exist two highly asymmetric Galactic HMXBs, we estimate the local rate of highly asymmetric GW190814-like events to be $\approx 0.5\,\mathrm{Gpc}^{-3}\mathrm{yr}^{-1}$, which is in broad agreement with GW observations.
\end{itemize}

As the ground-based GW detectors rapidly increase the sample of detected events, it will enable a better characterization of highly asymmetric mass-ratio BBH or NSBH systems \citep[e.g,][]{2024A&A...683A.144X}. It will allow us to understand the nature of individual events, as well as explore correlations in their mass, spin, and redshift distributions. Such studies could potentially distinguish between the different formation pathways proposed for the formation of such systems. Additionally, several ongoing or future surveys such as \emph{Gaia} \citep[e.g,][]{2023A&A...674A...1G, 2024NewAR..9801694E}, LSST \citep[e.g,][]{2023PASP..135j5002H}, Roman \citep[e.g,][]{2023arXiv230612514L}, eROSITA \citep[e.g,][]{2024A&A...682A..34M} will rapidly increase the sample of products of massive binary evolution in various stages. These systems provide a local anchors to understand various uncertainties in the evolution of massive stars and compact objects in binaries, and can be utilized as a bridge to understand the origin of GW sources.

\begin{acknowledgments}
N.S. would like to thank Thomas Callister, Amanda Farah, Nathan Smith and Gurtina Besla for useful discussions. M.R. acknowledges support from NASA (ATP: 95380NSSC24K0932) and NSF-AST-2510584. K.B. acknowledges support from NSF-AST-2510583 and the Falco-DeBenedetti Early Career Professorship. This work utilized the ElGato High Performance Computing cluster at the University of Arizona. The Theoretical Astrophysics Program (TAP) at the University of
Arizona provided resources to support this work. We respectfully acknowledge the University of Arizona is on the land and territories of Indigenous peoples. Today, Arizona is home to 22 federally recognized tribes, with Tucson being home to the O’odham and the Yaqui. The
university strives to build sustainable relationships with sovereign Native Nations and Indigenous communities through education offerings, partnerships, and community service.

This work has made use of the HMXB catalogue (\href{https://binary-revolution.github.io/HMXBwebcat/}{https://binary-revolution.github.io/HMXBwebcat/}) maintained by the Binary rEvolution team (\href{https://github.com/Binary-rEvolution}{https://github.com/Binary-rEvolution}).

This research has made use of data or software obtained from the Gravitational Wave Open Science Center (\href{gwosc.org}{gwosc.org}), a service of the LIGO Scientific Collaboration, the Virgo Collaboration, and KAGRA. This material is based upon work supported by NSF's LIGO Laboratory which is a major facility fully funded by the National Science Foundation, as well as the Science and Technology Facilities Council (STFC) of the United Kingdom, the Max-Planck-Society (MPS), and the State of Niedersachsen/Germany for support of the construction of Advanced LIGO and construction and operation of the GEO600 detector. Additional support for Advanced LIGO was provided by the Australian Research Council. Virgo is funded, through the European Gravitational Observatory (EGO), by the French Centre National de Recherche Scientifique (CNRS), the Italian Istituto Nazionale di Fisica Nucleare (INFN) and the Dutch Nikhef, with contributions by institutions from Belgium, Germany, Greece, Hungary, Ireland, Japan, Monaco, Poland, Portugal, Spain. KAGRA is supported by Ministry of Education, Culture, Sports, Science and Technology (MEXT), Japan Society for the Promotion of Science (JSPS) in Japan; National Research Foundation (NRF) and Ministry of Science and ICT (MSIT) in Korea; Academia Sinica (AS) and National Science and Technology Council (NSTC) in Taiwan.
\end{acknowledgments}

\section{Software}

This work made use of the following software packages: \texttt{astropy} \citep{astropy:2013, astropy:2018, astropy:2022}, \texttt{Jupyter} \citep{2007CSE.....9c..21P, kluyver2016jupyter}, \texttt{matplotlib} \citep{Hunter:2007}, \texttt{numpy} \citep{numpy}, \texttt{pandas} \citep{mckinney-proc-scipy-2010, pandas_17229934}, \texttt{python} \citep{python}, \texttt{scipy} \citep{2020SciPy-NMeth, scipy_17101542}, \texttt{Bilby} \citep{bilby_paper, bilby_paper_2, Bilby_17314023}, and \texttt{corner.py} \citep{corner-Foreman-Mackey-2016, corner.py_14209694}.

This research has made use of the Astrophysics Data System, funded by NASA under Cooperative Agreement 80NSSC21M00561.

This work uses Modules for Experiments in Stellar Astrophysics \citep[MESA][]{2011ApJS..192....3P,2013ApJS..208....4P,2015ApJS..220...15P,2018ApJS..234...34P,2019ApJS..243...10P,2023ApJS..265...15J}.

Software citation information aggregated using \texttt{\href{https://www.tomwagg.com/software-citation-station/}{The Software Citation Station}} \citep{software-citation-station-paper, software-citation-station-zenodo}.

\appendix

\section{Additional details about the MESA setup}

The MESA EOS is a blend of the OPAL \citep{Rogers2002}, SCVH \citep{Saumon1995}, FreeEOS \citep{Irwin2004}, HELM \citep{Timmes2000}, PC \citep{Potekhin2010}, and Skye \citep{Jermyn2021} EOSes. Radiative opacities are primarily from OPAL \citep{Iglesias1993, Iglesias1996}, with low-temperature data from \citet{Ferguson2005} and the high-temperature, Compton-scattering dominated regime by \citet{Poutanen2017}. Electron conduction opacities are from \citet{Cassisi2007} and \citet{Blouin2020}. Nuclear reaction rates are from JINA REACLIB \citep{Cyburt2010}, NACRE \citep{Angulo1999} and additional tabulated weak reaction rates \citet{Fuller1985, Oda1994, Langanke2000}. Screening is included via the prescription of \citet{Chugunov2007}. Thermal neutrino loss rates are from \citet{Itoh1996}. Roche lobe radii in binary systems are computed using the fit of \citet{1983ApJ...268..368E}.

\section{Resolution Tests} \label{resolution}

\begin{figure}
\gridline{\fig{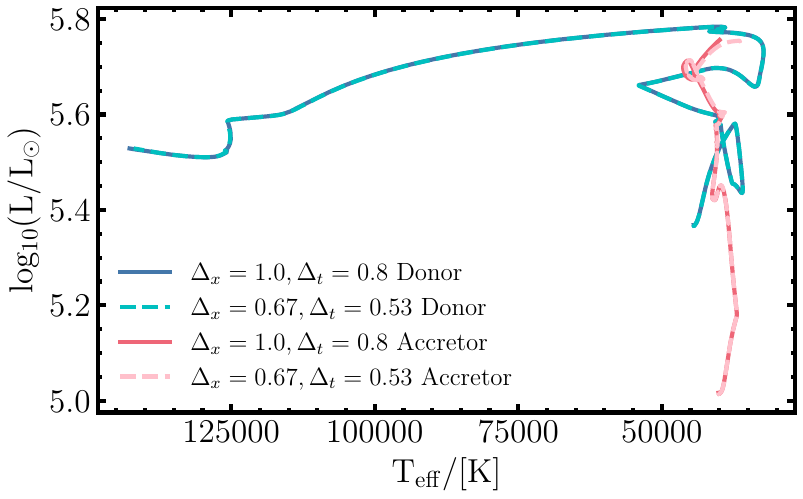}{0.47\textwidth}{(a)}
\fig{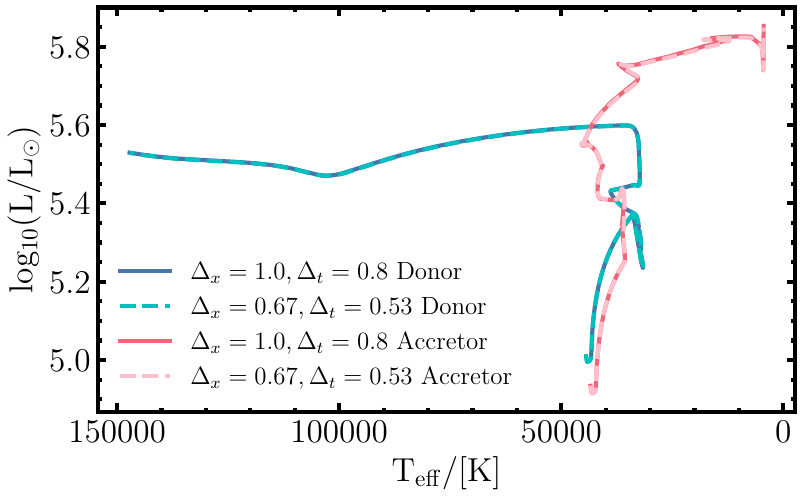}{0.47\textwidth}{(b)}}
\caption{(a) The evolution of the accretor and donor star, modeling the evolutionary history of the HMXB 4U 1700-37 on the HR diagram for two different choices of resolution. (b) The evolution of the accretor and donor star, modeling the formation of GW190814 on the HR diagram for two different choices of resolution.} \label{fig:resolution_test}
\end{figure}

To ensure that our results are robust to the choice of spatial and
temporal resolution in \textsc{MESA}, we
repeated our fiducial binary evolution calculations by increasing the number of mesh points, i.e. decreasing \texttt{mesh\_delta\_coeff} and \texttt{mesh\_time\_coeff} by a factor of $2/3$. In Fig \ref{fig:resolution_test}, we show that the evolution of the donor and accretor stars on the HR diagram for our fiducial models to explain the evolutionary history of the HMXB 4U 1700-37 (left panel) and a GW190814-like system (right panel). The blue and red solid colors represent the default resolution for the donor and accretor, where we have $\mathrm{\texttt{mesh\_delta\_coeff}} \equiv \Delta_x = 1$ and $\mathrm{\texttt{mesh\_time\_coeff}} \equiv \Delta_t = 0.8$. The cyan and pink dashed lines represent higher resolution models, where we decrease both $\Delta_x$ and $\Delta_t$ by a factor of $2/3$. We find that the evolutionary tracks of both the donor and accretor stars are nearly identical for both choices of resolution, demonstrating that our results are robust and there is convergence in the numerical treatment of binary evolution in \textsc{MESA}. The number of zones in our accretor models for the HMXB 4U 1700-37 are $\approx 1600$ and $\approx 2400$ for the fiducial and high resolution runs, while the number of timesteps are $\approx 67250$ and $\approx 68600$, respectively. The accretor models for the formation of a GW190814-like system have $\approx 1700$ and $\approx 2500$ zones for the fiducial and high resolution runs, while the number of timesteps are $\approx 14500$ and $\approx 17500$, respectively.

\section{Variations in Initial Conditions} \label{grid}

\begin{figure}[htbp]
    \centering
    \includegraphics[width=0.6\textwidth]{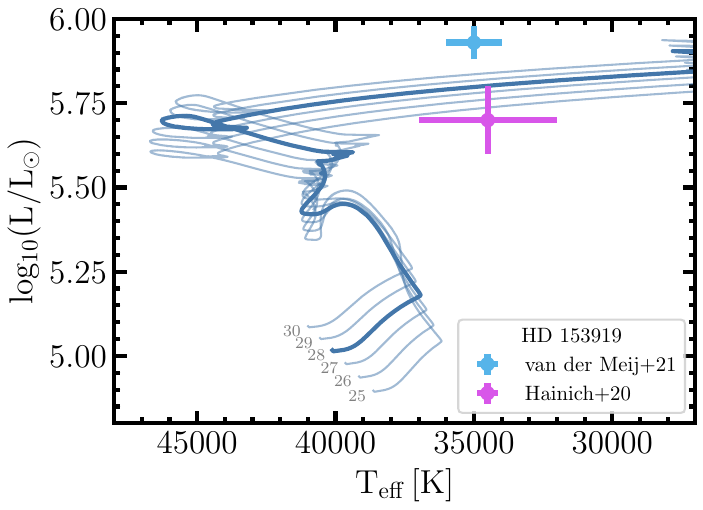}
    \caption{The evolution of accretors on the HR diagram. The dark blue curve highlights the fiducial model for the accretor in the formation of the HMXB 4U 1700-37/ HD 153919, while lighter shades represent variations in the mass of the accretor (denoted in gray in solar masses at the beginning of their evolution).}
    \label{fig:grid}
\end{figure}

To demonstrate that our binary evolution models for the formation of the HMXB 4U 1700-37/ HD 153919 are robust against small variations in initial parameters, we compute a small grid where we vary the initial mass of the accretor star from the value chosen for the fiducial model. Their evolution on the HR diagram is shown in Fig \ref{fig:grid}. All models follow a similar evolutionary track as the fiducial model. Depending on the mass of the accretor, the tracks shift up or down in luminosity. Additionally, lower mass accretors also spend more time on the Main Sequence post-SN, which may help address the discrepancy in the kinematic age of the system as discussed in Section \ref{age}. Our models can be further constrained with more observational studies of the surface properties and the distance to HD 153919, which will help resolve its exact position on the HR diagram.

\bibliography{xrb_gw}{}
\bibliographystyle{aasjournalv7}

\end{document}